\documentclass[12pt]{article}
\usepackage{amssymb,graphicx,amsmath,array,verbatim}

\usepackage{hyperref}

\makeatletter
\@addtoreset{equation}{section}

\renewcommand{\thefootnote}{\arabic{footnote}}
\makeatother

\newcommand{\N}{\mathcal{N}}

\newcommand {\beq}{\begin{eqnarray}}
\newcommand {\eeq}{\end{eqnarray}}
\def\p{\partial}
\newcommand{\NF}{N_{\rm F}}

\newcommand{\hs}[1]{\hspace{#1 mm}}
\newcommand{\bpm}{\begin{pmatrix}}
\newcommand{\epm}{\end{pmatrix}}

\newcommand{\R}{\mathbb{R}}
\newcommand{\C}{\mathbb{C}}

\newcommand{\D}{\mathcal D}

\newcommand{\ba}{\left( \begin{array}}
\newcommand{\ea}{\end{array} \right)}
\newcommand{\be}{\begin{equation}}
\newcommand{\ee}{\end{equation}}
\newcommand{\bea}{\begin{eqnarray}}
\newcommand{\eea}{\end{eqnarray}}
\newcommand{\beann}{\begin{eqnarray*}}
\newcommand{\eeann}{\end{eqnarray*}}

\newcommand{\Z}{\mathbb{Z}}

\usepackage[usenames]{color}

\begin{document}

\begin{titlepage}

\renewcommand{\thefootnote}{\fnsymbol{footnote}}

\begin{flushright}
\end{flushright}

\vspace{3em}

\begin{center}
 {\Large {\bf 
 2d Partition Function in $\Omega$-background\\
 and \vspace{.5em}\\ 
 Vortex/Instanton Correspondence
 }}

 \vskip5em

 {\sc Toshiaki Fujimori}$^1$,
 {\sc Taro Kimura}$^1$,
 {\sc Muneto Nitta}$^1$, and
 {\sc Keisuke Ohashi}$^2$


 \vskip2em

 $^1${\it 
 Department of Physics, and Research and Education Center for Natural
 Sciences,\\ Keio University, Hiyoshi 4-1-1, Kanagawa 223-8521, Japan
 }

 $^2${\it 
 Department of Physics, ``E. Fermi'', University of Pisa, Largo
 Pontecorvo, 3, 56127 Pisa, Italy\\
 INFN, Sezione di Pisa, Largo Pontecorvo, 3, 56127 Pisa, Italy
 }

 \vskip3em

\end{center}

 \vskip2em

\begin{abstract}
 We derive the exact vortex partition function 
 in 2d $\mathcal{N}=(2,2)$ gauge theory on the $\Omega$-background,
 applying the localization scheme in the Higgs phase.
 We show that the partition function at a finite $\Omega$-deformation parameter $\epsilon$ 
 satisfies a system of differential equations, 
 which can be interpreted as a quantized version of 
 the twisted $F$-term equations characterizing the SUSY vacua.
 Using the differential equations derived in this paper, 
 we show the correspondence between the partition function 
 of the two-dimensional vortex string worldsheet theory and 
 the Nekrasov partition function at the root of Higgs branch of the four-dimensional $\mathcal{N}=2$ theory 
 with two $\Omega$-deformation parameters $(\epsilon_1, \epsilon_2)$.
\end{abstract}

\end{titlepage}

\tableofcontents
\newpage

\section{Introduction}\label{sec:intro}

Recent progress on supersymmetric gauge theories allows us to perform
path integrals in an explicit way through the localization scheme.
Such exact results provide us with a number of non-trivial checks of
gauge theory dualities, e.g. gauge/gravity duality, mirror symmetry, Seiberg duality.
After the seminal work by Pestun \cite{Pestun:2007rz}, 
the localization scheme is widely applied to gauge theories on compact curved manifold, 
where the Infrared (IR) divergence is properly regularized
by the curvature of the spacetime. 
On the other hand, the exact computation on a non-compact spacetime needs 
another way to deal with the IR divergence
as in the earliest example of $\mathcal{N}=2$ supersymmetric gauge theories 
on the four-dimensional Euclidean spacetime~\cite{Nekrasov:2002qd}.
One possible approach for the non-compact case is to consider the $\Omega$-deformed theories, 
which can be naturally obtained through the standard dimensional reduction 
from the corresponding higher dimensional theory 
in a non-trivial supersymmetric background.
The exact partition function in 4d $\N=2$ gauge theories in the $\Omega$-background 
was computed in \cite{Nekrasov:2002qd} and 
it was shown that the partition function reproduces the Seiberg--Witten prepotential, 
which captures the low-energy dynamics of the theories.
Although, it cannot be applied to arbitrary dimensions,
it allows us to perform the path integral in several physically important classes of theories.

In this paper, we revisit the $\Omega$-background in two dimensions
and compute the exact vortex partition function~\cite{Shadchin:2006yz,Dimofte:2010tz,Yoshida:2011au,Bonelli:2011fq,Bonelli:2011wx,Dimofte:2011py,Fujimori:2012ab}, 
using the first-principles localization method in the Higgs phase, 
where the path integral localizes to BPS vortex configurations.
We first show the supersymmetry transformation in the $\Omega$-deformed two-dimensional theory, 
and then apply the localization method based on the (scalar) supersymmetry $\mathcal{Q}$ preserved by the $\Omega$-background.
To compute the one-loop determinant, 
we determine the bosonic and fermionic fluctuations around the BPS vortex configuration 
and count the short supermultiplets, which have non-trivial contribution to the partition function.
Combining the one-loop determinant and the classical contribution from a certain $\mathcal{Q}$-closed operator, 
we obtain both the explicit (combinatorial) form of the vortex partition function,
which can also be rewritten into an integral expression.  
As in the case of the four-dimensional theory, 
the whole partition function can be decomposed into the perturbative and the non-perturbative parts.
The latter part is the contribution from the BPS vortices, 
which play a role of instantons in two dimensions~
\cite{Hanany:2003hp,Auzzi:2003fs,Tong:2005un,Eto:2005yh,Eto:2006cx,Eto:2006pg,Shifman:2007ce,Tong:2008qd}.
We also show that the partition function does not simply factorize into two parts in the case involving fractional vortices.

Once the partition function in the $\Omega$-background is obtained, 
one can extract the non-trivial information for the low-energy dynamics of gauge theories
by taking the limit where the $\Omega$-background is turned off, $\epsilon \to 0$.
In addition to the role as the regulator, 
it is known that the $\Omega$-background parameter $\epsilon$ has another interesting
interpretation as the quantum deformation parameter~\cite{Nekrasov:2009rc,Nekrasov:2012xe,Nekrasov:2013xda}.
In this sense, the limit $\epsilon \to 0$ corresponds to the classical theory.
Indeed the Seiberg--Witten curve, which appears in this limit, has an
essential connection with the classical integrable systems.
At this moment, the Seiberg--Witten curve is just an algebraic curve,
which is characterized by an algebraic relation, $P(x,y)=0$.
Then, turning on the $\Omega$-background, the algebraic relation lifts
up to a differential equation $P(\hat{x},\hat{y})\cdot\psi=0$, which is called the quantum curve.
In our two-dimensional case, 
we show that the vortex partition function with finite $\epsilon$ satisfies a differential equation, 
where the corresponding differential operator is obtained by quantizing the twisted $F$-term equation, 
i.e. the algebraic relation characterizing the twisted superpotential.
This differential equation allows us to examine the behavior of the
partition function in the limit $\epsilon \to 0$ in an efficient way.

The $\N=(2,2)$ theory in two dimensions has various interesting
connection with $\N=2$ theory in four dimensions.
One of the important links between them is the correspondence between 
the BPS objects in two and four dimensions~\cite{Dorey:1998yh,Dorey:1999zk}.%
\footnote{
See~\cite{Aganagic:2013tta,Aganagic:2014oia,Aganagic:2015cta} for recent
progress along this direction, and also~\cite{Chen:2014rca,Chen:2015fta,Pan:2015hza}.
}
This correspondence has been interpreted as the relation between 
the 2d effective worldsheet theory on vortex strings 
and the parent 4d gauge theory at the root of Higgs branch \cite{Hanany:2004ea, Shifman:2004dr}. 
In this paper we apply the exact formula to obtain the vortex partition function 
in the effective worldsheet theory of vortex strings, 
called the Hanany--Tong model~\cite{Hanany:2003hp}.
We compare this vortex partition function 
with the four-dimensional Nekrasov partition function with two $\Omega$-deformation parameters $(\epsilon_1, \epsilon_2)$ 
and show the agreement of two and four-dimensional theories at the level of partition functions.

This paper is organized as follows.
In Sec.\,\ref{sec:part_func}, we first reconsider the vortex partition function in the $\Omega$-background.
Applying the localization technique in the Higgs phase, 
we derive the exact formula for the vortex partition function of Abelian $\N=(2,2)$ theories in two dimensions.
Deriving the integral expression for the partition function, 
we show that the mirror symmetry can be naturally observed by rewriting the integral.
We then generalize the result to the non-Abelian theory,
and show that it coincides with the hemisphere partition function~\cite{Sugishita:2013jca,Honda:2013uca,Hori:2013ika}.
In Sec.\,\ref{sec:diff}, we study the differential equation for the exact partition function.
It is derived by a contour shift of the integral formula for the partition function, 
which can be viewed as a ``quantized'' version of the twisted $F$-term equation.
In Sec.\,\ref{sec:HT}, we study the vortex worldsheet theory, 
the Hanany--Tong model, as an application of the formula obtained in this paper. 
By comparing the differential equations for the partition functions, 
we obtain the agreement of two- and four-dimensional partition functions at the root of Higgs branch 
with two $\Omega$-deformation parameters $(\epsilon_1, \epsilon_2)$.
In section~\ref{sec:summary}, we conclude this paper with some remarks.

\section{Vortex Partition Function}\label{sec:part_func}
In this section, we derive a formula for the vortex partition functions 
in 2d $\mathcal{N} = (2,2)$ supersymmetric theories in the $\Omega$-background. 
We apply the localization technique in the Higgs phase 
where the path integral localizes to BPS vortex configurations 
which are invariant under a certain symmetry transformation.
Since $\Omega$-deformed 2d theories can be obtained 
by means of the dimensional reduction from the corresponding 4d theories with a non-trivial metric, 
we first consider 4d $\mathcal{N}=1$ theories 
and apply the localization technique to compute the vortex partition function. 

\subsection{Omega Background}
The 4-dimensional geometry corresponding to 
the $\Omega$-background takes the form
\beq
ds^2 = \left| d z - i z (\epsilon dw + \bar{\epsilon} d \bar{w}) \right|^2 + |dw|^2, 
\eeq
where $z = x_1 + i x_2$, $w = x_3 + i x_4$ and 
$x_3$ and $x_4$ directions are periodic: 
$x_3 \sim x_3 + 2 \pi R_3$, $x_4 \sim x_4 + 2 \pi R_4$. 
The 2d $\Omega$-background can be obtained 
by performing the dimensional reduction 
along the periodic directions $x_3$ and $x_4$. 
If we take the limit $\epsilon \rightarrow 0$, 
the geometry reduces to $\R^2 \times T^2$ 
with the trivial metric and 
the resulting theory is the standard 2d $\mathcal{N}=(2,2)$ model
without any deformation. 
Let $e_i{}^\alpha~(\alpha=1,2,3,4)$ be the following basis for the vierbein 
\beq
dx^i (e_i{}^1+ie_i{}^2) = dz - i z (\epsilon dw + \bar{\epsilon} d \bar{w}), 
\hs{10}
dx^i (e_i{}^3 + i e_i{}^4) = dw.
\eeq
Then, the spin connection takes the form (see Appendix \ref{appendix:conventions} for our conventions of spinors)
\beq
\frac{1}{2} \omega_i^{\alpha \beta} \sigma_{\alpha \beta} dx^i ~=~  
\frac{1}{2} \omega_i^{\alpha \beta} \bar{\sigma}_{\alpha \beta} dx^i ~=~ 
\frac{1}{2} ( \epsilon dw + \bar{\epsilon} d \bar{w}) \sigma_{3}. 
\eeq

The supersymmetry transformation in this geometry is parameterized by 
the parallel spinors $\varepsilon$, 
$\bar{\varepsilon}$ which satisfy the following equations
\beq
\D_i \varepsilon = \left( \p_i + \frac{i}{2} \omega_i^{\alpha \beta} \sigma_{\alpha \beta} - i V_i \right) \varepsilon = 0, \hs{10} 
\D_i \bar{\varepsilon} = \left( \p_i + \frac{i}{2} \omega_i^{\alpha \beta} \bar{\sigma}_{\alpha \beta} + i V_i \right) \bar{\varepsilon} = 0,
\label{eq:spinor_eq}
\eeq
where $V_i$ is a background $R$-symmetry gauge field. 
In order to have supersymmetries in the $\Omega$-deformed geometry, 
the background $R$-symmetry gauge field should be appropriately turned on 
so that the parallel spinor equations \eqref{eq:spinor_eq} 
have non-trivial solutions. 
If we assume that $V_i$ is given by
\beq
V_i dx^i = -\frac{1}{2} ( \epsilon dw + \bar{\epsilon} d \bar{w} ), 
\eeq
the solutions of Eq.\,\eqref{eq:spinor_eq} 
satisfying the periodic boundary condition are given by
\beq
\varepsilon = \ba{c} 0 \\ \varepsilon_0 \ea, \hs{10}
\bar{\varepsilon} = \ba{c} \bar{\varepsilon}_0 \\ 0 \ea,
\eeq 
where $\varepsilon_0$ and $\bar{\varepsilon}_0$ are constants.
Note that for these spinors, the spinor bilinear $\varepsilon \bar{\sigma}^i \bar{\varepsilon}$ is
proportional to the Killing vector $\xi^i$ given by
\beq
\frac{1}{2 \varepsilon_0 \bar{\varepsilon}_0} (\varepsilon \bar{\sigma}^i \bar{\varepsilon}) \p_i 
~=~ \xi^i \p_i 
~\equiv~ \p_w + i \epsilon (z \p_z - \bar{z} \p_{\bar{z}}) .
\eeq

The field content of our model is the same 
as that of the standard $\mathcal{N}=1$ model in 4-dimensions. 
Furthermore, the supersymmetry variations of the fields 
with respect to the supercharge $\mathcal{Q} \equiv \varepsilon Q + \bar{\varepsilon} \bar{Q}$ 
also take the same form as those of $\mathcal{N}=1$ models 
if the background connections are 
appropriately incorporated into the covariant derivative. 
In the following, we first focus on the case of $U(1)$ gauge theory for simplicity. 
The supersymmetry transformation for a vector multiplet is given by
\begin{align}
\mathcal{Q} A_i &~=~ - i ( \bar{\lambda} \sigma_i \varepsilon - \bar{\varepsilon} \sigma_i \lambda ), 
&\mathcal{Q} \lambda &~=~- i F_{ij} \sigma^{ij} \varepsilon + D \varepsilon, \hs{10} 
\label{eq:vector_susy1} \\
\mathcal{Q} D &~=~ i ( \varepsilon \bar{\sigma}_i \D_i \bar{\lambda} + \D_i \lambda \bar{\sigma}_i \bar{\varepsilon} ), 
&\mathcal{Q} \bar{\lambda} &~=~ \phantom{-a} i F_{ij} \bar{\sigma}^{ij} \bar{\varepsilon} + D \bar{\varepsilon}, \hs{10} \label{eq:vector_susy2}
\end{align}
where the covariant derivative $\D_i$ contains 
the background $R$-symmetry gauge field 
\beq
\D_i \lambda = \left( \p_i + \frac{i}{2} \omega_i^{\alpha \beta} \sigma_{\alpha \beta} - i V_i \right) \lambda, \hs{10} 
\D_i \bar{\lambda} = \left( \p_i + \frac{i}{2} \omega_i^{\alpha \beta} \bar{\sigma}_{\alpha \beta} + i V_i \right) \bar{\lambda}.
\eeq
For a chiral multiplet, 
the supersymmetry transformation is given by
\begin{align}
\mathcal{Q}\phi &~=~ \sqrt{2} \varepsilon \psi, 
&\mathcal{Q} \bar{\phi} &~=~ \sqrt{2} \bar{\varepsilon} \bar{\psi}, \label{eq:chiral_susy1}\\
\mathcal{Q} \psi &~=~ \sqrt{2} ( i \D_i \phi \bar{\sigma}_i \bar{\varepsilon} + F \varepsilon ), 
&\mathcal{Q} \bar{\psi} &~=~ \sqrt{2} ( i \D_i \bar{\phi} \sigma_i \varepsilon + \bar{F} \bar{\varepsilon} ), \label{eq:chiral_susy2}\\ 
\mathcal{Q} F &~=~ \sqrt{2} ( i \D_i \psi \bar{\sigma}_i \bar{\varepsilon} + \sqrt{2} i \phi \bar{\varepsilon} \bar{\lambda} ), \hs{5}
&\mathcal{Q} \bar{F} &~=~ \sqrt{2} ( i \D_i \bar{\psi} \sigma_i \varepsilon + \sqrt{2} i \bar{\phi} \varepsilon \lambda ). \label{eq:chiral_susy3}
\end{align}
Again, the covariant derivative contains the background gauge fields $V_i$ for the $R$-symmetry, 
under which a chiral multiplet $(\phi, \psi, F)$ with $R$-charge $q$ transforms as
\beq
(\phi, \psi, F) ~\rightarrow~ e^{-i q \alpha} (\phi, e^{i \alpha} \psi, e^{2i \alpha} F).
\eeq 
In addition to the $R$-symmetry gauge field background, 
we can also introduce a flat gauge field background 
$\tilde{A}_i$ for each $U(1)$ subgroup of the Cartan part of the flavor symmetry
\beq
\tilde{A}_{i}^a dx^i ~=~ m^a dw + \bar{m}^a d\bar{w}.
\eeq
The complex parameters $m^a$ are called the twisted masses
in 2d $\mathcal{N}=(2,2)$ theories. 

Assuming that $\varepsilon_0$ and $\bar{\varepsilon}_0$ are commuting parameters, 
we can show that the square of the supersymmetry closes to 
the translation along the compact directions $P_w$, 
the rotation $J$ in the $z$-plane, 
the $R$-symmetry $R$
and the flavor symmetries $F_a$
\beq
\mathcal{Q}^2 = 4 \varepsilon_0 \bar{\varepsilon}_0 \left[ P_w + \epsilon \left( J - \frac{1}{2} R \right) + m^a F_a \right],
\label{eq:Q_square}
\eeq
where $P_w$ should be understood as the covariant derivative including the gauge connection.
If we redefine $\tilde{J} \equiv J - \frac{1}{2} R$ 
as a new generator for the rotation in the $z$-plane, 
the supercharges corresponding to the parameters 
$\varepsilon_0$ and $\bar{\varepsilon}_0$ 
can be regarded as scalars under $\tilde{J}$. 
Similarly, it is convenient to rewrite the spinor indices 
in terms of the new rotation generator $\tilde{J}$
to make the new rotational symmetry manifest
\beq
\ba{c} \lambda_1 \\ \lambda_2 \ea = \ba{c} \lambda_z \\ \lambda_0 \ea, \hs{10}
\ba{c} \bar{\lambda}^{\dot{1}} \\ \bar{\lambda}^{\dot{2}} \ea = \ba{c} \bar{\lambda}_0 \\ \bar{\lambda}_{\bar{z}} \ea, 
\eeq
\beq
\ba{c} \psi_1 \\ \psi_2 \ea = \ba{c} \psi_0 \\ \psi_{\bar{z}} \ea, \hs{10}
\ba{c} \bar{\psi}^{\dot{1}} \\ \bar{\psi}^{\dot{2}} \ea = \ba{c} \bar{\psi}_z \\ \bar{\psi}_{0} \ea,
\eeq
and the auxiliary field in a chiral multiplet $(F,\bar{F})$ becomes $(F_z, \bar{F}_{\bar{z}})$,  
where the quantities with the subscript 0 are scalars 
and $z,\,\bar{z}$ stand for $(1,0)$ and $(0,1)$-forms.
It is also convenient to use the following notation
\beq
A_z = \frac{1}{2} g^{ij} (e_i^1 - i e_i^2) A_j, \hs{10}
A_{\bar{z}} = \frac{1}{2} g^{ij} (e_i^1 + i e_i^2) A_j,
\eeq
\beq
A_{\xi} = \frac{1}{2} g^{ij} (e_i^3 - i e_i^4) A_j, \hs{10}
A_{\bar{\xi}} = \frac{1}{2} g^{ij} (e_i^3 + i e_i^4) A_j,
\eeq
and similarly for the field strength $F_{ij}$. Note that $A_{\xi}$ and $A_{\bar{\xi}}$ can also be viewed as the gauge connection in the directions of the Killing vector fields $\xi = \p_w + i \epsilon (z \p_z - \bar{z} \p_{\bar{z}})$ and $\bar{\xi} = \p_{\bar{w}} + i \bar{\epsilon} (z \p_z - \bar{z} \p_{\bar{z}})$, that is
\beq
A_{\xi} = \xi^i A_i, \hs{10} A_{\bar{\xi}} = \bar{\xi}^i A_i
\eeq
In terms of these new notations, 
the supersymmetry transformation rules for a vector multiplet
\eqref{eq:vector_susy1} and \eqref{eq:vector_susy2} can be rewritten as
\begin{align}
\mathcal{Q} A_z &~=~ - i \bar{\varepsilon}_0 \lambda_z, 
&\mathcal{Q} A_{\xi} &~=~ 0, \label{eq:vec_susy1} \\
\mathcal{Q} A_{\bar{z}} &~=~ - i \varepsilon_0 \bar{\lambda}_{\bar{z}}, 
&\mathcal{Q} A_{\bar{\xi}} &~=~ i (\bar{\varepsilon}_0 \lambda_0 - \varepsilon_0 \bar{\lambda}_0) , \label{eq:vec_susy2} \\ 
\mathcal{Q} \lambda_z &~=~ - 4 \varepsilon_0 F_{\xi z},
&\mathcal{Q} \lambda_0 &~=~ \varepsilon_0 ( 2 F_{z \bar{z}} + 2 F_{\xi \bar{\xi}} + D ), \label{eq:vec_susy3} \\
\mathcal{Q} \bar{\lambda}_{\bar{z}} &~=~ - 4 \bar{\varepsilon}_0 F_{\xi \bar{z}}
&\mathcal{Q} \bar{\lambda}_0 &~=~ \bar{\varepsilon}_0 ( 2 F_{z \bar{z}} - 2 F_{\xi \bar{\xi}} + D ), \label{eq:vec_susy4} \\
\mathcal{Q} D &~=~ 2 i \D_{\xi} ( \bar{\varepsilon}_0 \lambda_0 + \varepsilon_0 \bar{\lambda}_0 ) - 2 \mathcal{Q} F_{z \bar{z}}, 
&\mathcal{Q} F_{z \bar{z}} &~=~ i (\bar{\varepsilon}_0 \p_{\bar{z}} \lambda_z - \varepsilon_0 \p_z \bar{\lambda}_{\bar{z}}), \label{eq:vec_susy5}
\end{align}
and the supersymmetry transformation rules for a chiral multiplet 
\eqref{eq:chiral_susy1}-\eqref{eq:chiral_susy3} become
\begin{align}
\mathcal{Q} \phi &~=~ \hs{5} \sqrt{2} \varepsilon_0 \psi_0, 
&\mathcal{Q} \psi_0 &~=~ \hs{5} 2 \sqrt{2} i \bar{\varepsilon}_0 \D_\xi \phi, \\
\mathcal{Q} \bar{\phi} &~=~ - \sqrt{2} \bar{\varepsilon}_0 \bar{\psi}_0, 
&\mathcal{Q} \bar{\psi}_0 &~=~ - 2 \sqrt{2} i \varepsilon_0 \D_\xi \bar{\phi}, \\
\mathcal{Q} F_{\bar{z}} &~=~ 2 \bar{\varepsilon}_0 ( \sqrt{2} i \D_\xi \psi_{\bar{z}} - \sqrt{2} i \D_{\bar{z}} \psi_0 - i \phi \bar{\lambda}_{\bar{z}} ), 
&\mathcal{Q} \psi_{\bar{z}} &~=~ \sqrt{2} \left( 2 i \bar{\varepsilon}_0 \D_{\bar{z}} \phi + \varepsilon_0 F_{\bar{z}} \right), \\
\mathcal{Q} \bar{F}_{z} &~=~ 2 \varepsilon_0 ( \sqrt{2} i \D_\xi \bar{\psi}_{z} + \sqrt{2} i \D_{z} \bar{\psi}_0 + i \bar{\phi} \lambda_{z} ),  
&\mathcal{Q} \bar{\psi}_{z} &~=~ \sqrt{2} ( 2 i \varepsilon_0 \D_{z} \bar{\phi} + \bar{\varepsilon}_0 \bar{F}_{z} ), \label{eq:chi_susy4}
\end{align} 
where $\D_{\xi},~\D_{\bar{\xi}}$ are 
the covariant derivatives containing all the connections. 
For example, for the fermionic field in the $a$-th flavor, 
$\D_{\xi}$ is given by
\beq
\D_{\xi} \psi_{\bar{z}}^a = \left[ \xi^i \p_i + i A_{\xi} - i \epsilon + i \left(m_a + \frac{\epsilon q_a}{2} \right) \right]\psi_{\bar{z}}^a.
\eeq
In the following, we set $q_a=0$ 
since the $R$-charge, which appears in the combination 
$m_a+\epsilon q_a/2$, can always be absorbed\footnote{
If we set $q=0$ for all chiral multiplets 
in the presence of a superpotential $W$,  
the twisted masses have to be assigned so that 
each term in $W$ has the total twisted mass 
$2 \times \frac{\epsilon}{2} = \epsilon$ 
since the total mass and the $R$-charge of $W$ should have $m=0,~q=2$ in the original notation.} 
into the twisted mass $m_a$.
The $R$-charge dependence can be restored by shifting the twisted mass 
$m_a \rightarrow m_a+\epsilon q_a/2$. 

\subsection{$\mathcal{Q}$-exact Action and Saddle Points}
The vortex partition function can be defined by a path integral of the form 
\beq
Z = \int [\D \varphi] \, \exp \left( - \mathcal{Q} V + \mathcal{I} \right).
\label{eq:path_integral}
\eeq
Here and in the following, 
$\mathcal{Q}$ denotes the supercharge 
with $\varepsilon_0 = \bar{\varepsilon}_0 = \frac{1}{2}$.
$V$ and $\mathcal{I}$ are certain quantities satisfying
\beq
\mathcal{Q}^2 V = \mathcal{Q} \mathcal{I} = 0.
\eeq
Then, we can show that $Z$ is invariant 
under any deformation $V \rightarrow V + \delta V$ 
\beq
\delta Z = - \int [ \mathcal{D} \varphi ] \, (\mathcal{Q} \delta V) \exp \left( - \mathcal{Q} V + \mathcal{I} \right) 
= - \int [ \mathcal{D} \varphi ] \, \mathcal{Q} \Big[ \delta V \exp \left( - \mathcal{Q} V + \mathcal{I} \right) \Big] = 0.
\eeq
This property enables us to exactly evaluate the path integral $Z$ 
by using the deformation $V \rightarrow t V$ 
and applying the saddle point method in the limit $t \rightarrow \infty$. 

Schematically, the $\mathcal{Q}$-exact part $\mathcal{Q} V$ 
can be expanded around its saddle points in terms of bosonic and fermionic 
fluctuations $\Phi$ and $\Psi$: 
\beq
\mathcal{Q} V = \Phi^\dagger \Delta_B \Phi + \Psi^\dagger \Delta_F \Psi + \mbox{higher order terms},
\label{eq:action_fluctuation}
\eeq
where $\Delta_B$ and $\Delta_F$ are certain differential operators 
appearing in the linearized equations of motion around the saddle point configuration. 
The saddle point analysis implies that 
the vortex partition function can be written 
as a sum over the saddle points of $\mathcal{Q}V$ 
\beq
Z ~= \sum \bigg[ \exp(\mathcal{I}) \, \frac{\det(\Delta_F)}{\det(\Delta_B)} \ \bigg]_{\text{saddle points}}.
\label{eq:saddle_point}
\eeq
Therefore, the exact vortex partition function $Z$ can be obtained
by computing the contributions from each saddle points: 
the value of the $\mathcal{Q}$-closed operator $\mathcal{I}$ and 
the functional determinants of $\Delta_B$ and $\Delta_F$. 

To discuss the explicit form of $\mathcal{Q} V$ and its saddle points, 
we restrict ourselves to the Abelian model with $N$ chiral multiplets with electric charge $+1$. 
Let us consider the $\mathcal{Q}$-exact part generated from $V$ of the form 
\beq
V &=& \int d^4 x \bigg[ \frac{1}{g^2} \left( \lambda_z \overline{\mathcal{Q} \lambda_z} + \lambda_0 \overline{\mathcal{Q} \lambda_0} + \bar{\lambda}_0 \overline{\mathcal{Q} \bar{\lambda}_0} + \bar{\lambda}_{\bar{z}} \overline{\mathcal{Q} \bar{\lambda}_{\bar{z}}} \right) \notag \\
&& \hs{10} + ~\psi_0^a \overline{\mathcal{Q} \psi_0^a} + \psi_{\bar{z}}^a \overline{\mathcal{Q} \psi_{\bar{z}}^a} + \bar{\psi}_z^a \overline{\mathcal{Q} \bar{\psi}_z^a} + \bar{\psi}_0^a \overline{\mathcal{Q} \bar{\psi}_0^a} + i (\lambda_0 + \bar{\lambda}_0) (|\phi^a|^2 - r) \bigg], \label{eq:V}
\eeq
where $g$ and $r$ are constants corresponding 
to the gauge coupling constant and Fayet--Iliopoulos (FI) parameter. 
Note that the vortex partition function do not depend on such parameters 
contained in the $\mathcal{Q}$-exact part. 
For Eq.\,\eqref{eq:V}, $\mathcal{Q} V$ takes the form
\beq
\mathcal{Q} V = S_{4d} - 2 i r  \int d^4 x F_{z \bar{z}} + \mbox{\{total derivative terms\}},
\eeq
where $S_{4d}$ is the following action 
which reduces to that of the standard $\mathcal{N}=1$ supersymmetric model 
in the limit $\epsilon \rightarrow 0$: 
\beq
S_{4d} &=& \int d^4 x \bigg[ \frac{1}{4g^2} F_{ij} F^{ij} + \frac{1}{2g^2} D^2 - \frac{i}{2g^2} \bar{\lambda} \sigma^i \overset{\leftrightarrow}{\D}_i \lambda + \D_i \phi_a \D^j \bar{\phi}_a+ |F_a|^2 \notag \\
&& \hs{10} -~ \frac{i}{2} \bar{\psi}_a \sigma^i \overset{\leftrightarrow}{\D}_i \psi_a - \sqrt{2} i ( \phi_a \bar{\psi}_a \bar{\lambda} + \bar{\phi}_a \psi_a \lambda ) + i D (|\phi_a|^2 - r) \bigg] . \label{eq:4d_action}
\eeq
Note that the other terms in $\mathcal{Q} V$ are 
topological and total derivative terms. 
After integrating out the auxiliary fields $F_a$ and $D$, 
the bosonic part of $\mathcal{Q} V$ takes the form
\beq
(\mathcal{Q} V)_{\rm{bosonic}} &=& \int d^4x \bigg[ \frac{1}{2g^2} \left\{ \left| 2 i F_{z \bar{z}} + e^2(|\phi_a|^2 - r) \right|^2 + \left| F_{\xi \bar{\xi}} \right|^2 + 8|F_{\xi z}|^2 + 8|F_{\xi \bar{z}}|^2 \right\} \notag \\
&& \hs{10} +~ 4 |\D_{\bar{z}} \phi_a|^2 + 2 \big|\D_\xi \phi_a \big|^2 + 2 \big|\D_{\xi} \bar{\phi}_a \big|^2 \bigg].
\eeq
This completed square form implies that
the saddle points of $\mathcal{Q} V$ 
are the solutions of the following BPS equations
\beq
&\D_{\bar{z}} \phi_a \,=\, 0, \hs{5} 2 i F_{z \bar{z}} + g^2 (|\phi_a|^2 - r) \,=\, 0,& \label{eq:BPS1} \\
&F_{\xi \bar{\xi}} \,=\, F_{\xi z} \,=\, F_{\xi \bar{z}} \,=\, \D_\xi \phi_a \,=\, \D_{\xi} \bar{\phi}_a \,=\, 0.& \label{eq:BPS2}
\eeq 
The first two equations do not depend on the deformation parameter $\epsilon$ 
and can be identified with the BPS vortex equations 
in the corresponding $\mathcal{N}=(2,2)$ model without any deformation.  
The solution to Eqs.\,\eqref{eq:BPS1} can be formally written as
\beq
A_{\bar{z}} = - \frac{i}{2} \p_{\bar{z}} \omega, \hs{10}
\phi_a = \sqrt{r} \, e^{-\frac{1}{2} \omega} h_a(z),  
\label{eq:votex_sol}
\eeq
where $h_a(z)$ are arbitrary holomorphic polynomials of $z$. 
The profile function $\omega$ is the solution of the following equation \cite{Eto:2005yh,Eto:2006pg} 
\beq
\p_z \p_{\bar{z}} \omega = \frac{g^2 r}{2} ( 1 - H_0 H_0^\dagger e^{-\omega} ),
\label{eq:master}
\eeq
where $H_0(z) \equiv (h_1,\,h_2,\,\cdots,\,h_N)$ and 
the boundary condition is $\lim_{|z| \rightarrow \infty} \omega = \log (H_0 H_0^\dagger)$ .

In the undeformed massless theory $(m_a=\epsilon=0)$, 
each component of $H_0(z)$ can be an arbitrary polynomial, 
whose coefficients are regarded as the moduli parameters of the BPS vortex configurations. 
On the other hand, in the presence of 
the twisted mass and the omega deformation, 
Eqs.\,\eqref{eq:BPS2} tells us that $H_0(z)$ is restricted to 
the fixed points of the bosonic symmetry generated by 
\beq
{\mathcal{Q}}^2 = P_w + \epsilon (J - R/2) + m^a F_a.
\eeq 
Such configurations are specified by the vortex number $k \in \Z_{\geq 0}$ 
and the flavor label $a \in (1,2,\cdots,N)$ 
which specifies the vacuum at the spatial infinity.
For the $k$-vortex configuration in the $a$-th vacuum, 
the components of $H_0(z)$ are given by
\beq
h_a = z^k, \hs{10} h_b = 0~~(b \not = a),
\eeq
and Eqs.\,\eqref{eq:BPS2} is satisfied if 
\beq
A_{\xi} = - m_a - k \epsilon.
\eeq

In conclusion, the saddle point configuration specified by 
the vortex number $k$ and the vacuum label $a$ is given by
\beq
A_{\xi} = - m_a - k \epsilon, \hs{10}
A_{\bar{z}} = - \frac{i}{2} \p_{\bar{z}} \omega, \hs{10}
\phi_b = \left\{ 
\begin{array}{cc} 
\sqrt{r} \, e^{-\frac{1}{2} \omega} z^k & \mbox{for $b=a$} \\
0 & \mbox{for $b \not = a$} 
\end{array} \right..
\label{eq:saddle}
\eeq
Although the solution of Eq.\,\eqref{eq:master} is not known, 
the vortex partition function $Z$ can be determined
without the explicit use of the profile function $\omega$. 

\subsection{One-Loop Determinants}
Before specifying the $\mathcal{Q}$-closed operator $\mathcal{I}$ 
for the vortex partition function, 
let us first discuss the functional determinants in Eq.\,\eqref{eq:saddle_point}. 
For each saddle point configuration, 
the determinants can be calculated by finding the eigenmodes of 
the differential operators $\Delta_B$ and $\Delta_F$.
(see Appendix \ref{appendix:determinants} 
for the details of the computations). 

Since the saddle points are BPS configurations 
preserving the supersymmetry, 
the bosonic and fermionic fluctuations around them form supermultiplets. 
Generic supermultiplets do not contribute to the ratio of the determinants 
since they consist of one bosonic mode with eigenvalue 
$\Delta_B \doteq m_+ m_-$ and two fermionic modes with $\Delta_F \doteq m_{\pm}$.

Besides such generic supermultiplets, 
there exist short supermultiplets 
which consist of one bosonic and one fermionic modes, 
so that they have non-trivial contributions to the determinants. 
As shown in the Appendix \ref{appendix:determinants}, 
their eigenvalues of $\Delta_B$ and $\Delta_F$ are related 
to those of the covariant derivative $\D_\xi$ in the following way:
\beq
\Delta_B \doteq  - \D_{\xi} \D_{\bar{\xi}}, \hs{10} 
\Delta_F \doteq -i \D_{\bar{\xi}}. 
\eeq
Therefore, the ratio of the determinants can be obtained from 
the determinant of $\D_{\bar{\xi}}$ restricted to the short supermultiplets
\beq
\frac{\det \Delta_F}{\det \Delta_B} ~\propto~ \frac{1}{\det (-i \D_{\xi}) } \bigg|_{\rm{short}}. 
\label{eq:det_red}
\eeq
As shown in Appendix \ref{appendix:determinants}, 
the bosonic parts of such short multiplets are given by 
the solutions of the linearized version of the BPS vortex equations \eqref{eq:BPS1} in the undeformed theory. 
They can be obtained by linearizing 
the general form of the vortex solution Eq.\,\eqref{eq:votex_sol} 
around the saddle point configuration
\beq
\delta A_{\xi} = 0, \hs{10} 
\delta A_{\bar z} = -\frac{i}{2} \p_{\bar{z}} \delta \omega, \hs{10}
\delta \phi_b = \sqrt{r} e^{-\frac{1}{2} \omega} \left[ \delta h_b(z) -\frac{1}{2} \delta \omega \, h_{b}(z) \right] 
,& \label{eq:linear_sol}
\eeq
where $\delta h_b(z)$ are arbitrary holomorphic functions 
and once they are given, 
the function $\delta \omega$ can be uniquely determined 
through the linearized version of Eq.\,\eqref{eq:master} (see Eq.\,\eqref{eq:linear-master}). 

Although $\delta h_b(z)$ are arbitrary,
not all degrees of freedom contained in $\delta h_b(z)$ are physical
since the solution Eq.\,\eqref{eq:linear_sol} is invariant 
under the following infinite dimensional transformation: 
\beq
\delta h_b(z) \rightarrow \delta h_b(z) + v(z) h_b(z), \hs{10} 
\delta \omega \rightarrow \delta \omega + 2 v(z),
\label{eq:v-transf}
\eeq
where $v(z)$ is an arbitrary polynomial. 
Since $h_a(z) = z^k$ and $h_b(z) = 0~(b \not = a)$ 
for the $k$-vortex background in the $a$-th vacuum, 
only $\delta h_a(z)$ varies under this transformation. 
In order to find out the physical degrees of freedom, 
it is convenient to expand $\delta h_b(z)$ and 
fix the redundancy Eq.\,\eqref{eq:v-transf} by truncating the sum in $h_a(z)$ in the following way:
\beq
\delta h_a(z) = \sum_{l=0}^{k-1} c_{a, \, l} \, z^l, \hs{10}
\delta h_b(z) = \sum_{l=0}^{\infty} c_{b, \, l} \, z^l~~(b \not = a).
\eeq
where the coefficients $c_{b. l}$ are periodic functions of $x_3$ and $x_4$, 
so that they can be further expanded into the Kaluza-Klein modes
\beq
c_{b, \, l}(x_3, x_4)= \sum_{\mathbf{n}} c_{b, \, l}^{\mathbf{n}} \exp i \left( \frac{n_3 x_3}{R_3} + \frac{n_4 x_4}{R_4} \right), \hs{10} 
\mathbf{n} \equiv (n_3, \, n_4)
\eeq

There is a caveat here. 
It is well-known that the modes with $l \geq k-1~(b \not = a)$ are non-normalizable in the $k$-vortex background. 
In general, if there exists a non-normalizable mode, 
the spectrum is continuous and the boson-fermion cancellation argument may not be applicable.
However, we can show that there exists 
a continuous deformation of the $\mathcal{Q}$-exact part of the action 
under which all the modes become normalizable (see Appendix \ref{appendix:normalizability}). 
Therefore, there is no need to care about the normalizability of the eigenmodes
and the determinant can be calculated by using Eq.\,\eqref{eq:det_red}. 

The eigenvalue of $- i \D_{\xi}$ can be easily read off 
by noting that its action on $\Delta \phi_b$ reduces to the following action on $c_{b, \, l}^{\mathbf{n}} $:
\beq
- i \D_{\xi} \phi_b &\rightarrow& 
\lambda_{b, \, l}^{\mathbf{n}} \, c_{b, \, l}^{\mathbf{n}} 
~\equiv~ 
\left[ - \frac{i}{2} \left( \frac{n_3}{R_3} - i \frac{n_4}{R_4} \right) + m_b-m_a + \epsilon (l-k) \right] c_{b, \, l}^{\mathbf{n}}. 
\eeq 
The functional determinants can be obtained 
by taking the product of the eigenvalues 
$\lambda_{b, \, l}^{\mathbf{n}}$ for all the physical eigenmodes
\beq
\left[ \frac{\det \Delta_F}{\det \Delta_B} \right]_{4d,\,a,\,k} = ~ \prod_{\mathbf{n}} 
\left[ \left( \prod_{l=0}^{k-1} \frac{\Lambda_0}{\ \lambda_{a, \, l}^{\mathbf{n}}} \right) 
\prod_{b \not = a} \prod_{l=0}^\infty\frac{\Lambda_0}{\ \lambda_{b, \, l}^{\mathbf{n}}} \right],
\label{eq:det}
\eeq
where we have introduced a scale parameter $\Lambda_0$ 
so that the ratio of the determinants is dimensionless.
Now let us focus on the 2d limit $R_3,\,R_4 \rightarrow 0$, 
in which all the Kaluza Klein modes decouple 
and their contributions become trivial. 
Then, the one-loop determinant in two dimension can be obtained 
by ignoring the modes with $(n_3,n_4) \not= 0$
\beq
\left[ \frac{\det \Delta_F}{\det \Delta_B} \right]_{2d,\,a,\,k} = ~ \frac{(-1)^k}{k!} \left(\frac{\Lambda_0}{\epsilon} \right)^{N \left( \frac{m_a}{\epsilon} + k \right)} \prod_{b \not = a}\sqrt{\frac{\Lambda_0}{2\pi\epsilon}} \Gamma \left( \frac{m_b-m_a-k\epsilon}{\epsilon} \right),
\label{eq:determinant}
\eeq
where we have assumed $\sum_{b=1}^N m_b = 0$ 
(this can always be done by shifting the scalar field $\sigma$ in the 2d vector multiplet)
and the infinite products are reduced to the gamma functions by using the zeta function regularization\footnote{
The Hurwitz zeta function $\zeta(s,z) = \sum_{l=0}^\infty (z+l)^{-s}$ satisfies 
\beq
\zeta(0,z) = - z + \frac{1}{2} , \hs{10} 
\lim_{s \rightarrow 0} \frac{\p}{\p s} \zeta(s,z) = \log \frac{\Gamma(z)}{\sqrt{2\pi}}. \notag
\eeq}
\beq
\prod_{l=0}^\infty \frac{y}{z + l} = \exp \left[ \lim_{s \rightarrow 0} \frac{\p}{\p s} \Big\{ y^s \zeta \left( s, z \right) \Big\} \right] = y^{-z} \sqrt{\frac{y}{2\pi}} \Gamma(z).
\label{eq:zeta_ref}
\eeq

\subsection{$\mathcal{Q}$-closed Operator}
Let us next turn to the $\mathcal{Q}$-closed part $\mathcal{I}$. 
In the following, we will work in two dimensions unless otherwise stated. 
In order to write down $\mathcal{I}$, 
it is convenient to use the twisted chiral superfield for the vector multiplet,
which is expressed in terms of the anti-commuting Grassmann coordinates
$\theta_z$ and $\bar{\theta}_{\bar{z}}$ as\footnote{
Although this twisted chiral superfield $\Sigma$ 
is different from the one used in the standard $\mathcal{N}=(2,2)$ models, 
the difference vanishes when it is evaluated at the saddle points.}
\beq
\Sigma = \sigma + \frac{1}{2} \lambda_z \theta_{\bar{z}} + \frac{1}{2} \bar{\lambda}_{\bar{z}} \bar{\theta}_{z} + i F_{z \bar{z}} \theta_{\bar{z}} \bar{\theta}_{z},
\eeq
where $\sigma$ is the complex scalar field obtained from $A_w$ by the dimensional reduction. 
The supercharge which generates the transformation 
Eqs.\,\eqref{eq:vec_susy1}-\eqref{eq:vec_susy5} is now expressed as 
the following differential operator acting on twisted chiral superfields:
\beq
\mathcal{Q} = \theta_{\bar{z}} \frac{\p}{\p z} + \bar{\theta}_{z} \frac{\p}{\p \bar{z}} 
+ \epsilon \left( z \frac{\p}{\p \theta_{\bar{z}}} - \bar{z} \frac{\p}{\p \bar{\theta}_{z}} \right).
\eeq
With the identification $\theta_{\bar{z}} = d z,~\bar{\theta}_z=d\bar{z}$, 
this differential operator can be identified with 
the equivariant exterior derivative with the equivariant parameter $\epsilon$ for the $SO(2)$ rotation on the $z$-plane. 

Let $\widetilde{W}$ be a twisted superpotential, 
i.e. a holomorphic function of twisted chiral superfields.
Since the supersymmetry transformation $\mathcal{Q}$ acts on $\widetilde{W}$ as a total derivative, 
we can obtain a $\mathcal{Q}$-closed operator by integrating $\widetilde{W}$ over the superspace:
\beq
\mathcal{I} = 2 \int d^2 x \, d \bar{\theta}_{\bar{z}} \, d \theta_z \, \widetilde{W}(\Sigma). 
\eeq

Let $\tau_0$ be the following combination of 
the theta angle and the FI parameter: 
\beq
\tau_0 \equiv \frac{\theta}{2\pi} + i r. 
\eeq
The $\mathcal{Q}$-closed operator corresponding to the vortex partition function 
can be obtained from the standard tree-level twisted superpotential $\widetilde{W}= i \tau_0 \Sigma$
by promoting the constant parameter $\tau_0$ to a background twisted chiral superfield $T$:
\beq
\widetilde{W} = i T(\alpha) \Sigma,
\eeq
where we have assumed that $T$ is 
a function of the following ``equivariantly closed form'':
\beq
\alpha = \epsilon \left( \theta_{\bar{z}} \bar{\theta}_z + \epsilon |z|^2 \right), \hs{10} \mathcal{Q} \alpha = 0. 
\label{eq:equivariant}
\eeq
In addition, we also assume that  $T$ is a function satisfying
\beq
T(0) = \tau_0, \hs{10} \lim_{x \rightarrow \infty} T(x) = 0.
\label{eq:T_assump}
\eeq
For instance, $T(\alpha) = \tau_0 \, e^{-\alpha}$ is a typical example. 
As long as these conditions are satisfied, 
the explicit form of $T(\alpha)$ is not essential
since different choices of $T$ lead to $\mathcal{Q}$-closed operators 
which differ only by $\mathcal{Q}$-exact terms. 
The twisted superpotential preserves the supersymmetry 
since the background twisted chiral superfield 
$T(\alpha)$ is $\mathcal{Q}$-closed. 
For this choice of the twisted superpotential, 
the corresponding $\mathcal{Q}$-closed operator $\mathcal{I}$ is given by
\beq
\mathcal{I} = \int d^2 z \, 2 i \Big[ \epsilon T'(|\epsilon z|^2)  \sigma + i T(|\epsilon z|^2)  F_{z \bar{z}} \Big] 
= \int d^2 z \, 2 i \epsilon T'(|\epsilon z|^2) \Big[ \sigma + i \epsilon (z A_z - \bar{z} A_{\bar{z}}) \Big],
\label{eq:A_Qclosed}
\eeq
where we have used $\lim_{x \rightarrow \infty} T(x) = 0$. 
This quantity can be obtained 
from the following gauge invariant quantity in the 4d theory:
\beq
\mathcal{I}_{4d} &=& \frac{1}{(2\pi)^2 R_3 R_4} \int d^4 x \, 2 i \epsilon T'(|\epsilon z|^2) A_{\xi}.
\eeq
Therefore, at the saddle point Eq.\,\eqref{eq:saddle}, $\mathcal{I}$ takes the value 
\beq
\mathcal{I} \, \big|_{a,\,k} ~=~ \frac{2\pi}{\epsilon} \int_0^\infty d|\epsilon z|^2  \, i T'(|\epsilon z|^2) (-m_a - k \epsilon) ~=~ 2 \pi i \left( \frac{m_a}{\epsilon} + k \right) \tau_0.
\label{eq:Q_closed}
\eeq
where we have used $T(0) = \tau_0$. 

\paragraph{$Z(\tau_0)$ as a generating function \\}
The supersymmetry transformation rules \eqref{eq:vec_susy1}-\eqref{eq:vec_susy5} imply that 
in the limit $\epsilon \rightarrow 0$, the scalar field $\sigma$ becomes a $\mathcal{Q}$-closed operator 
whose difference at two arbitrary points is $\mathcal{Q}$-exact
\beq
\sigma |_{z=z_1} - \sigma |_{z=z_2} =~
\frac{1}{2} \int_{z_2}^{z_1} \mathcal{Q} (\lambda_z dz + \lambda_{\bar{z}} d \bar{z}) + \mathcal{O}(\epsilon).
\eeq
Thus, we can ignore the position dependence of $\sigma$ in the limit $\epsilon \rightarrow 0$, 
so that the leading order term in $\mathcal{I}$ is given by
\beq
\mathcal{I} = \int d^2 z \, 2 i T'(\epsilon|z|^2) \Big[ \sigma + \mathcal{O}(\epsilon) \Big] \sim - \frac{2\pi i \tau_0}{\epsilon} \Big[ \sigma + \mathcal{O}(\epsilon) \Big],
\label{eq:uptoQexact}
\eeq
where $\sim$ denotes the equality up to $\mathcal{Q}$-exact terms.
Since $\mathcal{Q}$-exact terms do not contribute to the path integral,  
Eq.\,\eqref{eq:uptoQexact} implies that the expectation value of $\sigma$ can be calculated
from $Z(\tau_0)$ in the following way\footnote{
Note that the FI term in $\mathcal{Q} V$ does not contribute to Eq.\,\eqref{eq:sigmaVEV}
since $Z$ does not depend on the parameters contained in the $\mathcal{Q}$-exact part.}
\beq
\langle \sigma \rangle 
~=~ \lim_{\epsilon \rightarrow 0} \left( - \frac{\epsilon}{2\pi i} \frac{\p}{\p \tau_0} \right) \log Z.
\label{eq:sigmaVEV}
\eeq
In the limit $\epsilon \rightarrow 0$, the twisted superpotential reduces to
the standard one $\widetilde{W} = i \tau_0 \Sigma$, 
and hence the action $-\mathcal{Q}V + \mathcal{I}$ becomes that of
the conventional $\mathcal{N}=(2,2)$ model with the topological $\theta$ term 
\beq
- \mathcal{Q} V + \mathcal{I} ~=~ - S_{2d} - \frac{i \theta}{2\pi} \int F + \{\mbox{total derivative}\}, 
\label{eq:2d_action}
\eeq
where $S_{2d}$ is the 2d action obtained from Eq.\,\eqref{eq:4d_action} by the dimensional reduction. 
Therefore, $\langle \sigma \rangle$ is the expectation value in the standard $\mathcal{N}=(2,2)$ model. 
It can also be obtained from the on-shell effective twisted superpotential: 
\beq
\langle \sigma \rangle = - i \p_{\tau_0} \widetilde{W}_{\rm eff}.
\eeq 
Comparing with Eq.\,\eqref{eq:sigmaVEV}, 
we find that the on-shell effective twisted superpotential
and $Z$ are related by
\beq
\widetilde{W}_{\rm eff}(\tau_0) = -\lim_{\epsilon \rightarrow 0} \frac{\epsilon}{2\pi} \log Z(\tau_0). 
\eeq
The off-shell effective twisted superpotential for $\sigma$ can be obtained from $\widetilde{W}_{\rm eff}(\tau_0)$
by the integrating-in procedure, i.e. the Legendre transformation.  
We will use it in the next section to check that our vortex partition function 
correctly reproduces the known results in $\mathcal{N}=(2,2)$ models.

\subsection{Vortex Partition Function and Integral Representations} 
Having obtained the one-loop determinant Eq.\,\eqref{eq:determinant} and 
the $\mathcal{Q}$-closed part Eq.\,\eqref{eq:Q_closed}, 
we can obtain the vortex partition function 
from the saddle point formula Eq.\,\eqref{eq:saddle_point}  
by summing over the saddle points. 
Since $z$-plane is non-compact, the vacuum at spatial infinity has to be fixed 
and hence the vortex partition function is defined for each vacuum. 
Summing over the vortex number $k$ in the $a$-th vacuum, 
we obtain the following form of the vortex partition function:
\beq
Z_{a} ~=~ \sum_{k=0}^\infty Z_{a,k} ~=~ \exp \left( \frac{2\pi i m_a \tau}{\epsilon} \right) 
\sum_{k=0}^\infty e^{2 \pi i k \tau} \frac{(-1)^k}{k!} \prod_{b \not = a} \Gamma \left( \frac{m_b-m_a-k\epsilon}{\epsilon} \right). 
\label{eq:VP_N}
\eeq
where we have ignored the overall factor $(\Lambda_0/\epsilon)^{\frac{N-1}{2}}$
and the UV parameter $\tau_0$ is replaced with renormalized parameter $\tau$ defined at $\mu=\epsilon$ 
\beq
2\pi i \tau = 2 \pi i \tau_0 + \log \frac{\Lambda_0^N}{\epsilon^N}. 
\eeq 
It is worth pointing out that the ratio of $Z_{a,k}$ and $Z_{a,0}$
agrees with the following finite dimensional integral on the vortex moduli space 
\beq
\frac{Z_{a,k}}{Z_{a,0}} ~=~ e^{2 \pi i k \tau} \int_{\mathcal{M}_{a,k}} \exp \left( - S_{{\rm eff},a,k} \right) 
~=~ e^{2 \pi i k \tau} \prod_{b =1}^{N} \prod_{j=1}^k \frac{\epsilon}{m_b-m_a-j\epsilon}, 
\eeq
where $S_{{\rm eff},a,k}$ is the $\mathcal{Q}$-exact vortex effective action.
This implies that $Z_{a,k}~(k=0,1,\cdots)$ have a common irrational part,
and hence the vortex partition function Eq.\,\eqref{eq:VP_N} factorizes as
\beq
Z_a =Z_{a}^{\rm pert} \, Z_{a}^{\rm vort}, 
\label{eq:factorization}
\eeq
where the perturbative part $Z_{a}^{\rm pert} \equiv Z_{a,0}$ and 
the vortex part $Z_{a}^{\rm vort} \equiv \sum_{k=0}^{\infty} Z_{a,k}/Z_{a,0}$ are given by
\begin{align}
{\textstyle 
Z_{a}^{\rm pert} = e^{\frac{2\pi i m_a \tau}{\epsilon}} \Gamma \left( \frac{m_b-m_a}{\epsilon} \right), \hs{10}
Z_{a}^{\rm vort} = {}_0F_{N-1} \left( \left\{ 1 - \frac{m_b-m_a}{\epsilon} \right\}_{b \not = a}, (-1)^N e^{2 \pi i \tau} \right)}.
\end{align}
The factorization Eq.\,\eqref{eq:factorization} 
is a general property of the vortex partition function 
except when there is contribution from the so-called fractional vortices (see Sec.\,\ref{sec:NA}). 

\begin{figure}[t]
\centering
\includegraphics[width=80mm]{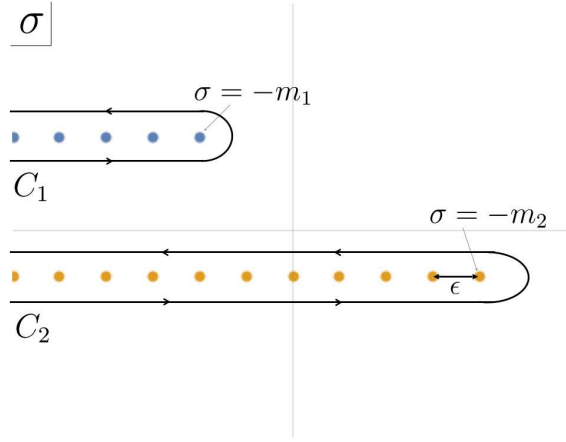}
\caption{The contours in the complex $\sigma$-plane ($\epsilon \in \R,\,\epsilon > 0$). 
The integration along $C_a$ gives the vortex partition function in the $a$-th vacuum. }
\label{fig:contour}
\end{figure}

We can also rewrite the result Eq.\,\eqref{eq:VP_N} into 
an integral form which will be more convenient in the discussion below. 
Since the gamma function $\Gamma(x)$ has poles at $x=-k \in \Z_{\leq 0}$ with residue $(-1)^k/k!$, 
the infinite sum over the vortex number $k$ can be rewritten
into the following contour integral along a path 
surrounding all the poles of the gamma function 
\beq
\sum_{k=0}^{\infty} \frac{(-1)^k}{k!} f(-k) = \int \frac{dx}{2\pi i} \Gamma(x) f(x).
\eeq
Applying this formula to Eq.\,\eqref{eq:VP_N} and 
changing the integration variable\footnote{
The new integration variable is denoted by $\sigma$ 
as the position of each pole is given by $\sigma|_{z=0}$ 
in the corresponding saddle point configuration:
\beq
\sigma \big|_{z=0} ~=~ \left[ -m_a - k \epsilon + \frac{1}{2} \epsilon (z \p_z + \bar{z} \p_{\bar{z}}) \omega \right]_{z=0}
~=~ - m_a - k \epsilon. \notag
\eeq}
$\sigma = \epsilon x - m_a$, 
we obtain the integral form of the vortex partition function
\beq
Z_a = \int_{C_a} \frac{d\sigma}{2\pi i \epsilon} \exp \left( - \frac{2\pi i \sigma \tau}{\epsilon} \right) \prod_{b=1}^N \Gamma \left( \frac{\sigma+m_b}{\epsilon} \right),
\label{eq:integral_cp}
\eeq 
Note that the integrand is independent of the vacuum label $a$,  
so that only the integration contour has the information on the choice of vacuum. 
The vortex partition function in the $a$-th vacuum can be obtained by 
choosing a contour which encloses all the poles of $a$-th Gamma function (see Fig.\,\ref{fig:contour}). 

\paragraph{A dual description \\}
As is well-known, our model has the mirror dual description 
in which the dynamical degrees of freedom are described by twisted chiral multiplets \cite{Hori:2000kt}. 
The vortex partition function can also be expressed 
in terms of the twisted superpotential in the dual theory.
In the case of $U(1)$ gauge theory with $N$ charged chiral multiplets, 
the integral form of the vortex partition function 
Eq.\,\eqref{eq:integral_cp} can be rewritten into the following form
\beq
Z_a = \int \frac{d\sigma}{2\pi i \epsilon} d y_1 \cdots d y_N \exp \left( - \frac{2\pi}{\epsilon} \widetilde{W} \right), 
\label{eq:integral_cp2}
\eeq
where $\widetilde{W}$ is the following twisted superpotential in the dual theory:
\beq
\widetilde{W} = i \sigma \tau  - \frac{1}{2\pi} \sum_{a=1}^N  \Big[ (\sigma+m_b)y_a - \epsilon e^{-y_a} \Big].
\eeq
Noting that each integral over $y_a$ gives the gamma function
\beq
\Gamma(x) = \int_{-\infty}^{\infty} dy  \exp \left( x y - e^y \right),
\eeq
we can easily see that this integral expression \eqref{eq:integral_cp2} reduces to Eq.\,\eqref{eq:integral_cp}. 
The existence of the integral representation of $Z$ 
in terms of the dual twisted superpotential implies that  
Eq.\,\eqref{eq:integral_cp2} can be directly obtained 
as a result of the localization computation in the $\Omega$-deformed mirror model.

\subsection{Generalization to Non-Abelian Gauge Theories}
\label{sec:NA}
Our result can be easily generalized to non-Abelian theories 
with arbitrary gauge groups and matter representations. 
Let us consider a model with gauge group $G$ (${\rm rank} \, G = r$)
and matter chiral multiplets in representation $\otimes_{a=1}^{\NF} R^a$. 
The saddle point equations in four dimensions are given by
\beq
&\D_{\bar{z}} \phi_a = 0, ~~~~~\, 2 i F^\alpha_{z \bar{z}} + g^2 ( \phi_a^\dagger T_{R^a}^\alpha \phi_a - r^\alpha) = 0, & \label{eq:NA_BPS1} \\
&F_{\xi \bar{\xi}} ~=~ F_{\xi z} ~=~ F_{\xi \bar{z}} ~=~ \D_\xi \phi_a ~=~ \D_{\xi} \bar{\phi}_a ~=~ 0,& \label{eq:NA_BPS2}
\eeq
where $T^\alpha_{R^a}~(\alpha=1,\cdots, {\rm dim} \, G)$ are the generators of the gauge group $G$ and 
$r^\alpha$ are FI parameters which are non-zero only for the central $U(1)$ parts of the gauge group.
As in the Abelian case, the first two equations can be identified with 
the BPS vortex equations in the trivial 2d plane 
\cite{Hanany:2003hp,Auzzi:2003fs,Tong:2005un,Eto:2005yh,Eto:2006cx,Eto:2006pg,Shifman:2007ce,Tong:2008qd}
and the other equations restrict the vortex configurations to 
be at the fixed points of the isometry $\mathcal{Q}^2$ on the moduli space. 

In the following, we restrict ourselves to the case where the saddle point configurations have no flat direction. 
Then we can show that each saddle point are specified by 
a set of flavor labels $\mathbf{a}=(a_1,a_2,\cdots,a_r)$,  
a linearly independent set of weight vectors 
$\vec{\boldsymbol{\mu}}=(\vec{\mu}_1,\vec{\mu}_2,\cdots,\vec{\mu}_r) \in (R^{a_1},R^{a_2},\cdots,R^{a_r})$
and a set of non-negative integers $\mathbf{k}=(k_1,k_2,\cdots,k_r)$. 
The flavor labels $\mathbf{a}$ and the weight vectors $\vec{\boldsymbol{\mu}}$ specify 
the non-zero components of the scalar fields\footnote{\noindent
The $D$-term condition requires that if $a_i = a_j$, the difference of the corresponding weight vectors 
cannot be on the root lattice $\vec{\mu}_i - \vec{\mu}_j \not \in \Delta_G$.
}
$\phi_{a}$, whereas the integers $\mathbf{k}$ denote their winding numbers. 
In general, the saddle point configurations take the form
\beq
A_{\bar{z}} = - \frac{i}{2} \vec{H} \cdot \p_{\bar{z}} \vec{\omega}, \hs{10} 
A_{\xi} = \vec{H} \cdot \vec{\sigma}_{\ast}, \hs{10}
\phi_{\rho}^a = \exp \left( - \frac{1}{2} \vec{\rho} \cdot \vec{\omega} \ \right) h_{\rho}^a(z),
\eeq
where $\vec{H}$ stands for the generators of the Cartan subalgebra $\mathfrak{t} \subset \mathfrak{g}$ and
$\vec{\omega}=(\omega_1,\omega_2,\cdots,\omega_r)$ are profile functions, 
whose explicit forms are not important. 
Each component of the scalar fields $\phi_{\rho}^a~(a=1,\cdots,\NF,~\vec{\rho} \in R^a)$ 
is determined by a holomorphic polynomial $h_{\rho}^a(z)$. 
For the saddle point specified by $\mathbf{a}, \vec{\boldsymbol{\mu}}$ and $\mathbf{k}$, 
the polynomials $h_{\rho}^a(z)$ are given by
\beq
h_{\rho}^a(z) = \left\{ 
\begin{array}{cl} 
z^{k_i} & (\mbox{$a=a_i,\,\vec{\rho}=\vec{\mu}_i$}) \\ 
0 & (\mbox{otherwise}) 
\end{array} \right..
\eeq 
The constants\footnote{
In two dimensions, $\vec{\sigma}_{\ast}$ specify the eigenvalues of the vector multiplet scalar field $\sigma$ at $z=0$.} 
$\vec{\sigma}_{\ast}=(\sigma_{01},\sigma_{02},\cdots,\sigma_{0r})$ can be determined from Eq.\,\eqref{eq:NA_BPS2},  
which reduces to the following condition: 
\beq
\vec{\mu}_i \cdot \vec{\sigma}_{\ast} = - m_{a_i} - \epsilon k_i \hs{5} (i=1,\cdots,r).
\label{eq:saddle_cond}
\eeq
Since the weight vectors $(\vec{\mu}_1,\vec{\mu}_2,\cdots,\vec{\mu}_r)$ are linearly independent, 
this system of equations uniquely determine $\vec{\sigma}_{\ast}$. 

As in the Abelian case, the computation of the one-loop determinants essentially reduces to 
the problem of finding solutions of the linearized BPS vortex equations.  
We can show that they are specified by a set of holomorphic functions $\delta h^a_{\rho}(z)$:
\beq
\delta \phi_{\rho}^a = e^{- \frac{1}{2} \vec{\rho} \cdot \vec{\omega}} \Big[  \delta h^a(z) + \delta \Omega \cdot  h^a \Big]_{\rho}, \hs{10} 
\delta A_{\bar{z}} = e^{-\frac{1}{2} \vec{H} \cdot \vec{\omega}} \, (i \p_{\bar{z}} \delta \Omega) \, e^{\frac{1}{2} \vec{H} \cdot \vec{\omega}}, 
\eeq
and $\delta A_\xi = 0$. Here, $\delta \Omega(z,\bar{z})$ is a function 
which takes value in the complexified Lie algebra $\mathfrak{g}^\C$ and 
in principle,  it can be determined once $\delta h^a_{\rho}(z)$ are given. 
In the 2d limit, each term in the Taylor expansion $ \delta h_{\rho}^a(z) = \sum c_{\rho,\,l}^a \, z^l$ 
has the eigenvalue $-i \D_{\xi} \doteq \vec{\rho} \cdot \vec{\sigma}_{\ast} + m_a + l \epsilon$, 
so that each component $\delta h_{\rho}^a$ gives 
the following contribution to $\det(-i \D_{\xi})^{-1}$: 
\beq
\frac{1}{\det(-i \D_{\xi})} \bigg|_{a,\rho} ~=~ 
\prod_{l=0}^\infty \frac{\Lambda_0}{\vec{\rho} \cdot \vec{\sigma}_{\ast} + m_a + l \epsilon} 
~=~ \left( \frac{\Lambda_0}{\epsilon} \right)^{-\frac{\vec{\rho} \cdot \vec{\sigma}_{\ast} + m_a}{\epsilon}} 
\Gamma \left( \frac{\vec{\rho} \cdot \vec{\sigma}_{\ast} + m_a}{\epsilon} \right).
\label{eq:naive}
\eeq
The total contribution can be obtained by taking the product over all the components $\delta h_{\rho}^a$ 
and eliminating the contributions from the unphysical modes generated 
by the infinitesimal complexified gauge transformation defined by
\beq
\delta h^a \sim \delta h^{a}(z) + v(z) \cdot h^{a}(z), \hs{5} \delta \Omega \sim \delta \Omega - v(z). 
\eeq
The transformation parameter $v(z)$ is 
an arbitrary element of the complexified Lie algebra $\mathfrak g^{\C}$, 
which can be decomposed into the diagonal and off-diagonal parts
\beq
v(z) = \vec{P}(z) \cdot \vec{H} + \sum_{\vec{\alpha} \in \Delta_G} Q_{\vec{\alpha}}(z) E_{\vec{\alpha}},
\eeq
where $\vec{P}(z)$ and $Q_{\vec{\alpha}}(z)$ are arbitrary polynomials. 

By using the Cartan part $\vec{P}(z) \cdot \vec{H}$, 
the fluctuations of the non-zero scalar fields can be fixed to finite order polynomials
\beq
\delta h_{\rho}^{a} = \sum_{l=0}^{k_i-1} c_{\rho, l}^{a} \, z^l \hs{10} (\mbox{for $a=a_i,~\vec{\rho}=\vec{\mu}_i$}). 
\eeq
Therefore, the contributions from 
the unphysical modes generated by the diagonal part $v(z)=\vec{P}(z) \cdot \vec{H}$
can be eliminated by replacing the infinite products Eq.\,\eqref{eq:naive} for $a=a_i, \, \vec{\rho} = \vec{\mu}_{i}$ 
with the finite ones
\beq
\prod_{l=0}^{k_i-1} \frac{\Lambda_0}{\vec{\mu}_{i} \cdot \vec{\sigma}_{\ast} + m_{a_i} + l \epsilon} = 
\left( - \frac{\Lambda_0}{\epsilon} \right)^{k_i} \frac{1}{k_i!},
\label{eq:unphysical1}
\eeq
where we have used Eq.\,\eqref{eq:saddle_cond}.
On the other hand, the off-diagonal part of the form $v(z) = z^l E_{\vec{\alpha}}$ 
generates the following unphysical mode:
\beq
\delta h_a = z^l E_{\vec{\alpha}} \cdot h_a. 
\label{eq:offdiagonal_unphys}
\eeq 
The eigenvalue of this unphysical fluctuation is $-i \D_{\xi} \doteq \vec{\alpha} \cdot \vec{\sigma}_{\ast} + l \epsilon$. 
Therefore, the total contribution from the off-diagonal part 
$\sum Q_{\vec{\alpha}}(z) E_{\vec{\alpha}}$ can be eliminated by dividing
$\det (-i\D_{\xi})^{-1}$ with\footnote{
By using the reflection formula for the gamma function,
the unphysical contribution can also be written as
\beq
\left[ \prod_{\alpha \in \Delta_G} \Gamma \left( \frac{\vec{\alpha} \cdot \vec{\sigma}_{\ast}}{\epsilon} \right) \right]^{-1} = \prod_{\vec{\alpha} > 0} \frac{\vec{\alpha} \cdot \vec{\sigma}_{\ast}}{\pi \epsilon} \sin \left( -\frac{\pi \vec{\alpha} \cdot \vec{\sigma}_{\ast}}{\epsilon} \right). \notag
\eeq
This is called the Sklyanin measure~\cite{Sklyanin:1984sb}.
}
\beq
\prod_{\alpha \in \Delta_G} \prod_{l=0}^\infty \frac{\Lambda_0}{\vec{\alpha} \cdot \vec{\sigma}_{\ast} + l \epsilon} 
= \prod_{\alpha \in \Delta_G} \Gamma \left( \frac{\vec{\alpha} \cdot \vec{\sigma}_{\ast}}{\epsilon} \right).
\label{eq:unphysical2}
\eeq

For the $\mathcal{Q}$-closed part $\mathcal{I}$, 
we consider an analogue of Eq.\,\eqref{eq:A_Qclosed},  
whose value at the saddle point is given by
\beq
\exp \left( \mathcal{I} \right) \big|_{\mbox{\footnotesize{saddle point}}} ~=~ 
\exp \left( - \frac{2\pi i \vec{\sigma}_{\ast} \cdot \vec{\tau}_0}{\epsilon} \right).
\label{eq:Q-closed_NA}
\eeq
Although the FI parameters and theta angle can be assigned only 
for each central $U(1)$ factor in the gauge group $G$, 
we have introduced auxiliary parameters $\vec{\tau}_0 =(\tau_{01}, \cdots, \tau_{0r})$ 
for all the $U(1)$ factors in the Cartan subgroup for later convenience. 

Combining Eq.\,\eqref{eq:Q-closed_NA} with the functional determinant Eq.\,\eqref{eq:naive} and 
taking into account the unphysical contributions Eq.\,\eqref{eq:unphysical1} and \eqref{eq:unphysical2}, 
we obtain the following contribution to the vortex partition function 
from the saddle point ($\mathbf{a}, \boldsymbol{\vec{\mu}}, \mathbf{k}$)
\beq
Z_{\mathbf{a},\boldsymbol{\mu},\mathbf{k}} = \exp \left( - \frac{2\pi i \vec{\sigma}_{\ast} \cdot \vec{\tau}}{\epsilon} \right) 
\left[ \prod_{\vec{\alpha} \in \Delta_G} \Gamma \left (\frac{\vec{\alpha} \cdot \vec{\sigma}_{\ast}}{\epsilon} \right) \right]^{-1} 
\left[ \prod_{i=1}^{r} \frac{(-1)^{k_i}}{k_i!} \right] 
{\prod_{a, \rho}}' \ \Gamma \left( \frac{\vec{\rho} \cdot \vec{\sigma}_{\ast} + m_a}{\epsilon} \right),
\label{eq:VP_general} 
\eeq
where ${\prod}'$ denotes 
the product over all the chiral multiplets excluding $(a, \vec{\rho}) =( a_i,\,\vec{\mu}_i)~(i=1,\cdots,r)$. 
Note that the UV scale has been absorbed into the renormalized coupling constants $\vec{\tau}$
\beq
\frac{2\pi i \vec{\sigma}_{\ast} \cdot \vec{\tau}}{\epsilon} ~=~ \frac{2\pi i \vec{\sigma}_{\ast} \cdot \vec{\tau}_0}{\epsilon} + \sum_{a=1}^{\NF} \sum_{\vec{\rho} \in R_a} \frac{\vec{\rho} \cdot \vec{\sigma}_{\ast}}{\epsilon} \log \left( \frac{\Lambda_0}{\epsilon} \right) .
\eeq
The vortex partition function in the vacuum specified by $(\mathbf{a}, \vec{\boldsymbol{\mu}})$
can be obtained by summing up Eq.\,\eqref{eq:VP_general} 
over the winding numbers: 
$Z_{\mathbf{a},\boldsymbol{\mu}} = \sum_{\mathbf{k}} Z_{\mathbf{a},\boldsymbol{\mu},\mathbf{k}}$.
One can easily see from Eq.\,\eqref{eq:VP_general} that 
$Z_{\mathbf{a},\boldsymbol{\mu}}$ can be rewritten into the following integral form: 
\beq
Z_{\mathbf{a},\boldsymbol{\mu}} = \int_{C_{\mathbf{a}, \boldsymbol{\mu}}} \prod_{i=1}^{r} \left( \frac{d \sigma_i}{2\pi i \epsilon} \right) \exp \left( -\frac{2\pi i \vec{\sigma} \cdot \vec{\tau}}{\epsilon} \right) \left[ \prod_{\vec{\alpha} \in G} \Gamma \left( \frac{\vec{\alpha} \cdot \vec{\sigma}}{\epsilon} \right) \right]^{-1} \prod_{a=1}^{\NF} \prod_{\vec{\rho} \in R_a} \Gamma \left( \frac{\vec{\rho} \cdot \vec{\sigma} + m_a}{\epsilon} \right),
\label{eq:integral_NA}
\eeq
where $C_{\mathbf{a}, \boldsymbol{\mu}}$ is a contour which encloses the set of poles located 
at $\vec{\sigma} = \vec{\sigma}_\ast$ satisfying the saddle point condition 
$\vec{\boldsymbol{\mu}} \cdot \vec{\sigma}_\ast = -\mathbf{m} - \epsilon \mathbf{k}$ 
with $\mathbf{k}=(k_1,\cdots,k_r) \in \Z^r_{\geq 0}$.

Let us remark that the expression \eqref{eq:integral_NA} is indeed
equivalent to the hemisphere partition function
formula~\cite{Sugishita:2013jca,Honda:2013uca,Hori:2013ika} with the
Neumann boundary condition applied to the chiral multiplets.
This agreement is reasonable because the two-dimensional Euclidean space
$\mathbf{R}^2$ is topologically equivalent to the hemisphere, where the
infinity in this case plays a role of the boundary.
From this point of view, it would be interesting to study a situation
corresponding to the Dirichlet boundary condition.
It shall be possible if we put physical degrees of freedom at infinity.

\paragraph{Fractional vortices \\}
Here, we briefly see that the factorization of property 
of the vortex partition function Eq.\,\eqref{eq:factorization}
is slightly modified when there exist the so-called fractional vortices \cite{Eto:2009bz}.
For the set of weight vectors $\boldsymbol{\mu}=(\vec{\mu}_1, \cdots, \vec{\mu}_r)$, 
let $\{\vec{\mu}_1^{\ast},\cdots,\vec{\mu}_r^{\ast}\}$ be the set of vectors such that
\beq
\vec{\mu}_i \cdot \vec{\mu}_j^{\ast} = \delta_{ij}.
\eeq
Once the solution to the saddle point condition Eq.\,\eqref{eq:saddle_cond} for $\mathbf{k}=0$ is given, 
those for other configurations can be written as
\beq
\vec{\sigma}_{\mathbf{a},\boldsymbol{\mu},\mathbf{k}} = 
\vec{\sigma}_{\mathbf{a},\boldsymbol{\mu},\mathbf{0}} - \epsilon \sum_{i=1}^r k_i \vec{\mu}_i^{\ast}.
\eeq
In other words, the difference 
$\vec{d}_{\mathbf{a},\boldsymbol{\mu},\mathbf{k}} \equiv \epsilon^{-1} (
\vec{\sigma}_{\mathbf{a},\boldsymbol{\mu},\mathbf{k}}-\vec{\sigma}_{\mathbf{a},\boldsymbol{\mu},\mathbf{0}})$
is on the lattice $\Lambda_{\boldsymbol{\mu}}^{\ast}$ generated by $\{\vec{\mu}_1^{\ast},\cdots,\vec{\mu}_r^{\ast}\}$, 
which contains the cocharactor lattice $\Lambda_{\rm cochar}$ as a sublattice
\beq
\Lambda_{\boldsymbol{\mu}}^{\ast} \equiv \Big\{ \vec{\lambda} \ \Big| \ \vec{\lambda} \cdot \vec{\mu}_i \in \Z,~1 \leq i \leq r \Big\} ~\supseteq~
\Lambda_{\rm cochar} \equiv \left\{ \vec{\lambda} \ \Big| \ \vec{\lambda} \cdot \vec{\rho} \in \Z, ~\forall \vec{\rho} \in \cup_a R_a \right\}.
\eeq
If there is a weight $\vec{\rho} \in \cup_a R_a$ such that $\vec{\rho} \cdot \vec{\mu}_i^{\ast} \not \in \Z$,
the lattice $\Lambda_{\boldsymbol{\mu}}^{\ast}$ contains $\Lambda_{\rm cochar}$ as a proper subgroup, 
i.e. $\Lambda_{\boldsymbol{\mu}}^{\ast} \not = \Lambda_{\rm cochar}$.
In such a case, one can show that
the classical vacuum specified by $\boldsymbol{\mu}$ has an unbroken discrete gauge symmetry. 
It has been known that in such a classical vacuum, 
there can be the so-called fractional vortices, whose magnetic fluxes are fractional numbers. 
For a saddle point with $\vec{d}_{\mathbf{a},\boldsymbol{\mu},\mathbf{k}} \not \in \Lambda_{\rm cochar}$, 
the magnetic flux $\int F \propto \vec{H} \cdot \vec{d}_{\mathbf{a},\boldsymbol{\mu},\mathbf{k}}$ has fractional eigenvalues, 
and hence there exist fractional vortices in such a saddle point configuration. 

In the previous section, 
we have seen in Eq.\,\eqref{eq:factorization} 
that the vortex partition function factorizes 
into the perturbative and vortex parts.
However, this is not always the case when there exist fractional vortices. 
This is because the ratio $Z_{\mathbf{a},\boldsymbol{\mu},\mathbf{k}}/Z_{\mathbf{a},\boldsymbol{\mu},\mathbf{0}}$ 
is not necessarily a rational function which can be written as a finite dimensional integral over the vortex moduli space
\beq
\frac{Z_{\mathbf{a},\boldsymbol{\mu},\mathbf{k}}}{Z_{\mathbf{a},\boldsymbol{\mu},\mathbf{0}}} \sim 
e^{2 \pi i k_i \, \vec{\mu}_i^{\ast} \cdot \vec{\tau}} 
\prod_{b,\rho} \frac{\Gamma \left( x_{b,\rho} - k_i \, \vec{\rho} \cdot \vec{\mu}_i^\ast \right)}{\Gamma \left( x_{b,\rho} \right)}, \hs{10} 
x_{b,\rho} \equiv \frac{\vec{\rho} \cdot \vec{\sigma}_{\mathbf{a},\boldsymbol{\mu},\mathbf{0}} + m_b}{\epsilon}.
\eeq
In such a case, the vortex partition function decomposes into several sectors 
labeled by the quotient $\Lambda_{\boldsymbol{\mu}}^{\ast}/\Lambda_{\rm cochar}$, 
each of which has a factorized form
\beq
Z_{\mathbf{a},\boldsymbol{\mu}} = 
\sum_{w} \,
Z_{\mathbf{a},\boldsymbol{\mu},p}^{\rm pert} \,
Z_{\mathbf{a},\boldsymbol{\mu},p}^{\rm vortex}, \hs{10} 
w \in \Lambda_{\boldsymbol{\mu}}^{\ast}/\Lambda_{\rm cochar}.
\eeq
We will see an example of fractional vortices in the next section. 

Let us comment on another realization of fractional vortices, which appear in the two-dimensional orbifold $\mathbb{C}/\mathbb{Z}_p$ plane~\cite{Kimura:2011wh} for which the vortex counting was studied in Ref.~\cite{Zhao:2011ke}.
This fractional vortex is different from that mentioned here, because the current one corresponds to orbifolding the target space.
Although these two are different, it would be interesting to study them with emphasis on the similarity.

\section{Differential Equation for the Vortex Partition Function}\label{sec:diff}
\subsection{$U(1)$ Gauge theories with $N$ Chiral multiplets}
In the previous section, we have obtained the integral expression for the vortex partition function. 
It is also convenient to characterize the vortex partition function 
in terms of linear differential equation which gives $Z_a(\tau)$ 
as certain linear combinations of its independent solutions.
\begin{figure}[t]
\centering
\includegraphics[width=70mm]{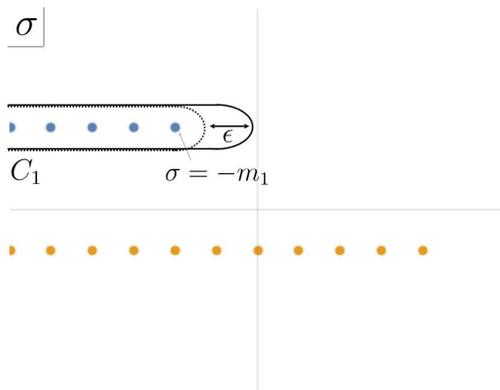}
\caption{The vortex partition function $Z_1$ is invariant under the shift of the contour $C_1$. }
\label{fig:shift}
\end{figure}

Let us first start with the simplest example of $U(1)$ gauge theory with $N$ charged chiral multiplets. 
The vortex partition function Eq.\,\eqref{eq:integral_cp} 
is invariant under an arbitrary deformation of the integration contour $C_a$ 
as long as it does not cross any pole of the integrand.
Shifting the contour by $\epsilon$ and 
changing the integration variable $\sigma \rightarrow \sigma + \epsilon$, 
we obtain
\beq
Z_a ~=~ \mbox{Eq.\,\eqref{eq:integral_cp}} ~=~ e^{-2\pi i \tau} \int_{C_a} \frac{d\sigma}{2\pi i \epsilon} \exp \left( - \frac{2\pi i \sigma \tau}{\epsilon} \right) \prod_{b=1}^N \frac{\sigma+m_a}{\epsilon} \Gamma \left( \frac{\sigma +m_b}{\epsilon} \right).
\label{eq:relation1}
\eeq
This relation can be rewritten into the following linear differential equation:
\beq
P\left( \hat \sigma \right) Z_{a} \equiv \left[ \prod_{b=1}^N ( \hat \sigma + m_b ) - \epsilon^N e^{2 \pi i \tau} \right] Z_{a} = 0, 
\label{eq:diffv}
\eeq 
where we have defined the differential operator $\hat{\sigma}$ by
\beq
\hat \sigma ~\equiv~ - \frac{\epsilon}{2\pi i} \p_{\tau}.
\eeq
We can see that $Z_a~(a=1,\cdots,N)$ in Eq.\,\eqref{eq:VP_N}
are the linearly independent solutions of this differential equation
for which $\epsilon \log Z_a$ is analytic around $\epsilon = 0$. 
This analyticity is an essential condition for the vortex partition function
since $Z_a$ should have the following behavior for small $\epsilon$: 
\beq
Z_a = \exp \left[ -\frac{2\pi}{\epsilon} \widetilde{W}_{a} + \mathcal O(1) \right].
\eeq
As we have seen in the previous section, 
$\widetilde{W}_{a}$ is the on-shell value of the twisted superpotential in the $a$-th vacuum, 
which is related to the VEV of $\sigma$ as
\beq
\langle \sigma \rangle_a = -\frac{2\pi}{\epsilon} \hat{\sigma} \widetilde{W}_a . 
\eeq
This implies that for small $\epsilon$, the operator $\hat{\sigma}$ is replaced by its expectation value
\beq
~ P\big( \hat \sigma \big) Z_a ~=~ P\big( \langle \sigma \rangle_a \big) Z_a + \mathcal O(\epsilon) ~=~ 0.
\eeq
Therefore, the $\epsilon \rightarrow 0$ limit of the differential equation \eqref{eq:diffv}
implies that  the expectation value of $\sigma$ in each vacuum agrees with
one of the roots of the polynomial $P(\sigma)$ given by
\beq
P(\sigma)= \prod_{a=1}^N(\sigma + m_a) - \Lambda^N. 
\eeq
Note that we have replaced the parameter $\tau(\epsilon)$ renormalized at $\mu = \epsilon$ 
with the $\mu$-independent scale $\Lambda$ defined by
\beq
\Lambda = \mu  \exp \left( \frac{2\pi i \tau(\mu)}{N} \right),  ~~~ \mbox{for arbitrary $\mu$}.
\eeq

\paragraph{The off-shell twisted superpotential \\}
Once the on-shell effective twisted superpotential $\widetilde{W}_a$ is obtained, 
the off-shell effective twisted superpotential $\widetilde{W}_{\rm eff}(\tau,\,\sigma)$ 
can also be obtained by the so-called integrating-in procedure
\beq
\widetilde{W}_{\rm eff}(\tau,\,\sigma) = i \sigma \tau(\mu) + \widetilde{W}_L(\sigma), 
\eeq
where $\widetilde{W}_L(\sigma)$ is the Legendre transform of the on-shell twisted superpotential 
\beq
\widetilde{W}_L(\sigma) = \left( \widetilde{W}_{\rm on}(\tau) - i \sigma \tau \right) \bigg|_{\tau = \tau(\sigma)}, \hs{10}
\tau(\sigma) = \frac{1}{2\pi i} \log \prod_{a=1}^N \frac{\sigma + m_a}{\mu}.
\eeq
From the effective twisted superpotential $\widetilde{W}_{\rm eff}(\tau,\,\sigma)$, 
the twisted $F$-term equation, 
which determines the expectation value of $\sigma$, 
can be obtained as
\beq
1 = \exp\left( 2\pi \p_\sigma \widetilde{W}_{\rm eff} \right) = e^{2\pi i \tau(\mu)} \exp \left( 2 \pi W_L'(\sigma) \right).
\label{eq:F}
\eeq
On the other hand, we have seen that the expectation values of $\sigma$ satisfies $P(\sigma)=0$. 
Comparing these two equations, we conclude that
\beq
\widetilde{W}_{\rm eff}(\tau,\sigma) = i \sigma \tau(\mu) - \frac{1}{2\pi} \sum_{a=1}^N (\sigma + m_a) \left( \log \frac{\sigma + m_a}{\mu} - 1 \right).
\label{eq:Wsigma}
\eeq
up to an additive constant which does not depend on $\sigma$. 
This agrees with the known result \cite{DAdda:1982eh, Cecotti:1992rm, Witten:1993yc, Morrison:1994fr}.

\subsection{Generalization to Non-Abelian Gauge Theories}
We have seen in the previous example that 
in the presence of the $\Omega$ deformation parameter $\epsilon$, 
the twisted $F$-term equation $P(\sigma)=0$ is promoted to 
the differential equation $P(\hat{\sigma})Z=0$.
In this sense, the differential equation $P(\sigma) Z = 0$ can be regarded as 
a ``quantized'' version of the twisted F-term equation. 
The $\Omega$ deformation parameter $\epsilon$ plays the role of the ``Planck constant'' 
characterizing the commutation relation
\beq
[ \hat{\sigma}, \hat{\tau} ] = i \frac{\epsilon}{2\pi}. 
\label{eq:quantization}
\eeq
This also holds in the model with arbitrary gauge group and matter content. 
In general, there is a differential equation 
for each elements of the cocharacter lattice $\Lambda_{\rm cochar}$, 
i.e. for each vector $\vec{\lambda}$ such that 
$\vec{\rho} \cdot \vec{\lambda} \in \Z,~\forall \vec{\rho} \in \cup_a R_a$. 
For each $\vec{\lambda} \in \Lambda_{\rm cochar}$, 
let $P_{\lambda}^{\pm}(\vec{\sigma})$ be 
the coprime polynomials of $\vec{\sigma}=(\sigma_1,\cdots,\sigma_r)$ 
defined by
\beq
\frac{P_{\lambda}^{+}(\vec{\sigma})}{P_{\lambda}^{-}(\vec{\sigma})} &=& 
\prod_{a=1}^{\NF} \prod_{\rho \in R_{a}} \epsilon^{\vec{\rho} \cdot \vec{\lambda}} \,
\Big( \frac{\vec{\rho} \cdot \vec{\sigma} + m_a}{\epsilon} \Big)_{\vec{\rho} \cdot \vec{\lambda}} ,
\eeq
where $(n)_m \equiv \Gamma(n+m)/\Gamma(n)$ denotes the Pochhammer symbol. 
From the invariance of Eq.\,\eqref{eq:integral_NA}  
under the shift of contour $C_{\mathbf{a}, \boldsymbol{\mu}}$ by $\epsilon \vec{\lambda}$, 
we find that the vortex partition function satisfies the following differential equation:
\beq
\bigg[ \Delta (\vec{\hat{\sigma}} + \epsilon \vec{\lambda}) P_{\lambda}^+({\vec{\hat{\sigma}}}) - \epsilon^{\vec{\rho}_t \cdot \vec{\lambda}} \exp \left\{ 2\pi i \vec{\lambda} \cdot (\vec{\tau}+\vec{\rho}_w) \right\} \Delta (\vec{\hat{\sigma}} - \epsilon \vec{\lambda}) P_{\lambda}^-({\vec{\hat{\sigma}}}) \bigg] Z = 0, 
\label{eq:NA_diffeq}
\eeq
where we have defined
\beq
\vec{\rho}_t = \sum_{a=1}^{\NF} \sum_{\vec{\rho} \in R_a} \vec{\rho}, \hs{10}
\vec{\rho}_w = \frac{1}{2} \sum_{\vec{\alpha} > 0} \vec{\alpha}, \hs{10}
\Delta(\vec{\sigma}) = \prod_{\vec{\alpha} > 0} \vec{\alpha} \cdot \vec{\sigma}.
\eeq
The vector $\vec{\hat{\sigma}} = (\hat{\sigma}_1, \cdots, \hat{\sigma}_r)$ denotes
the differential operators defined by
\beq
\hat{\sigma}_i = - \frac{\epsilon}{2\pi i} \frac{\p}{\p \tau_i}.
\eeq 
Note that not all vectors $\vec{\lambda} \in \Lambda_{\rm cochar}$ 
give independent differential equations 
but those for the basis $\vec{\lambda}_i~(i=1,\cdots,r)$ are independent.
In the limit $\epsilon \rightarrow 0$, 
the differential equation reduces to the following algebraic equation for $\vec{\sigma}$
\beq
\prod_{a=1}^{\NF} \prod_{\rho \in R_{a}} \Big\{ \vec{\rho} \cdot \vec{\sigma} + m_a \Big\}^{\vec{\rho} \cdot \vec{\lambda}} = 
\mu^{\vec{\rho}_t \cdot \vec{\lambda}} \exp \left[ 2\pi i \vec{\lambda} \cdot \left\{ \vec{\tau}(\mu)+\vec{\rho}_w \right\} \right], 
\label{eq:TF_general}
\eeq
where we have changed the renormalization scale from $\epsilon$ 
to a generic scale $\mu$ by using the following relation
\beq
2 \pi i \vec{\tau} = 2 \pi i \vec{\tau}(\mu) + \vec{\rho}_t \, \log \frac{\mu}{\epsilon}.
\eeq
Eq.\,\eqref{eq:TF_general} can be viewed as the ``classical limit''
of the differential equation Eq.\,\eqref{eq:NA_diffeq} and
as expected, it is the twisted $F$-term equation 
obtained from the following effective twisted superpotential:
\beq
\widetilde{W}_{\rm eff} = 2 \pi i \vec{\sigma} \cdot \vec{\tau}(\mu) 
+ \sum_{\alpha} f \left( \frac{\vec{\alpha} \cdot \vec{\sigma}}{\mu} \right) 
- \sum_{a=1}^{\NF} \sum_{\rho \in R_a} f \left( \frac{\vec{\rho} \cdot \vec{\sigma} + m_a}{\mu} \right) .
\eeq
where $f(x) \equiv x (\log x - 1)$. 

\paragraph{Fractional vortices \\} 
Here, we discuss an example of the fractional vortices in terms of the differential equation for $Z$. 
Let us consider the $U(1)$ gauge theory 
coupled to two chiral multiplets with charge $\rho_1=1$ and $\rho_2=q \in \Z$. 
For $\lambda=1$, the differential equation is given by
\beq
\bigg[ ( \hat \sigma + m_1 ) \prod_{j=0}^{q-1} (q \, \hat \sigma + m_2 + j \epsilon)- \epsilon^{q+1} e^{2 \pi i \tau} \bigg] Z = 0.
\label{eq:diff_ex}
\eeq
The general solution of this differential equation 
is the linear combination of the following hypergeometric functions
\beq
f_1 \,\, &=& \textstyle e^{2 \pi i \frac{m_1}{\epsilon} \tau} {}_0 F_q \left( \left\{ 1 + \frac{s-i}{q}  \right\}_{0 \leq i \leq q-1} ; -\frac{e^{2\pi i \tau}}{(-q)^q} \right) ,\\
f_{2,j} &=& \textstyle e^{2 \pi i \frac{m_2}{q \epsilon} \tau} {}_0 F_q \left( \left\{1 + \frac{j-i}{q} \right\}_{i \not = j}, 1 + \frac{j-s}{q} ; -\frac{e^{2\pi i \tau}}{(-q)^q} \right) e^{ 2\pi i  \frac{j}{q} \tau} , 
\label{eq:1_2_sol}
\eeq
where $s = \frac{q m_1-m_2}{\epsilon}$.  
The coefficients of the linear combination for $Z$ in each vacuum should be chosen 
so that $\epsilon \p_\tau \log Z$ is analytic around $\epsilon = 0$.
Two such linear combinations
\beq
 Z_1 = \Gamma \left( - s \right) f_1, \hs{10} 
 Z_2 = \sum_{j=0}^{q-1} \textstyle \Gamma \left( \frac{s-j}{q} \right) f_{2,j},
\eeq
agree with the following integral expression of $Z$ along the contours $C_1$ and $C_2$ 
enclosing the poles at $\sigma = - m_1 - \epsilon k$ and $(-m_2-\epsilon k)/q$: 
\beq
Z_a = \int_{C_a} \frac{d \sigma}{2\pi i} \, e^{- \frac{2 \pi i \sigma \tau}{\epsilon}} 
\Gamma \left( \frac{\sigma + m_1}{\epsilon} \right) \Gamma \left( \frac{q \, \sigma + m_2}{\epsilon} \right)
\eeq
Note that $Z_2$ consists of the $q$ sectors, each of which has the contribution $\exp(2 \pi i \tau j/q)$ from $j$ fractional vortices. 
Since the differential equation Eq.\,\eqref{eq:diff_ex} is invariant under the shift $\tau \rightarrow \tau+1$, 
other analytic solutions can be obtained 
by applying the shifts $\tau \rightarrow \tau + p ~ ( p = 1 , \cdots , q-1 )$ to $Z_2$
\beq
Z_{2, p} = \sum_{j=0}^{q-1} \textstyle e^{\frac{2 \pi i p}{q}} \Gamma \left( \frac{s-j}{q} \right) f_{2,j}. 
\eeq
As this examples shows, if there exist fractional vortices, 
the vortex partition function is not invariant 
under the shift of the theta angle $\theta \rightarrow \theta + 2\pi$.
This is the consequence of the fractionality of the magnetic flux.

\section{Vortex Counting on Vortex String Worldsheet}\label{sec:HT}
In this section, we discuss vortex strings in four dimensions 
whose worldsheet effective dynamics is described 
by a certain 2d model with $\mathcal{N}=(2,2)$ supersymmetry. 
It has been known that instanton-like vortices on the vortex string worldsheet 
can be identified with Yang--Mills instantons from the viewpoint of the parent 4d theory \cite{Hanany:2004ea,Eto:2004rz,Fujimori:2008ee}. 
By using the differential equation discussed in the previous section, 
we show that the vortex partition functions in the 2d vortex worldsheet theory 
agrees with the Nekrasov partition function 
at the so-called ``root of the Higgs branch'' in four dimensions. 

\begin{figure}[htb]
\begin{minipage}{0.5\hsize}
\centering
\includegraphics[width=7cm]{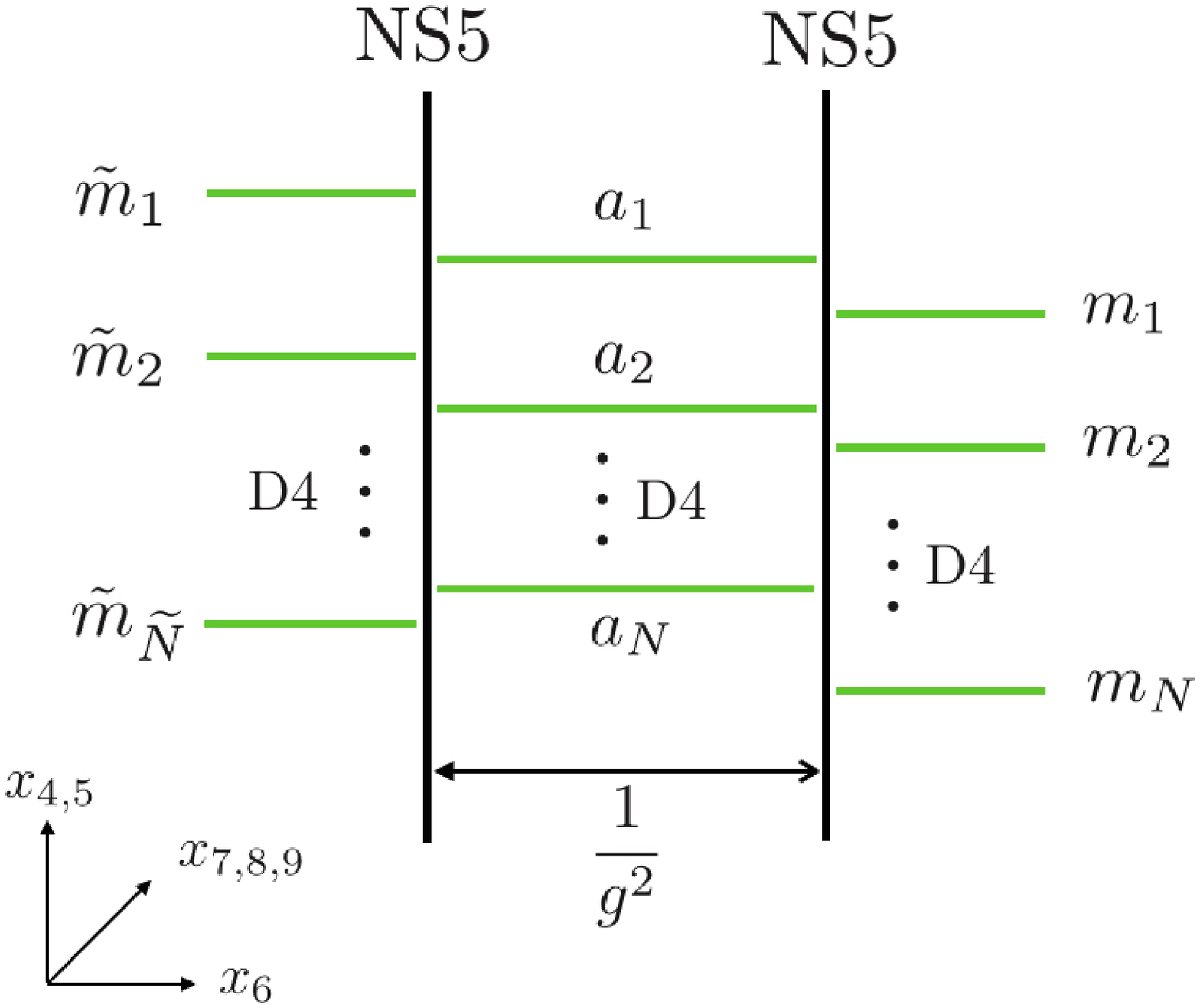}
\caption{The Hanany--Witten brane setup}
\label{fig:HWbrane}
\end{minipage}
\begin{minipage}{0.5\hsize}
\centering
\includegraphics[width=7cm]{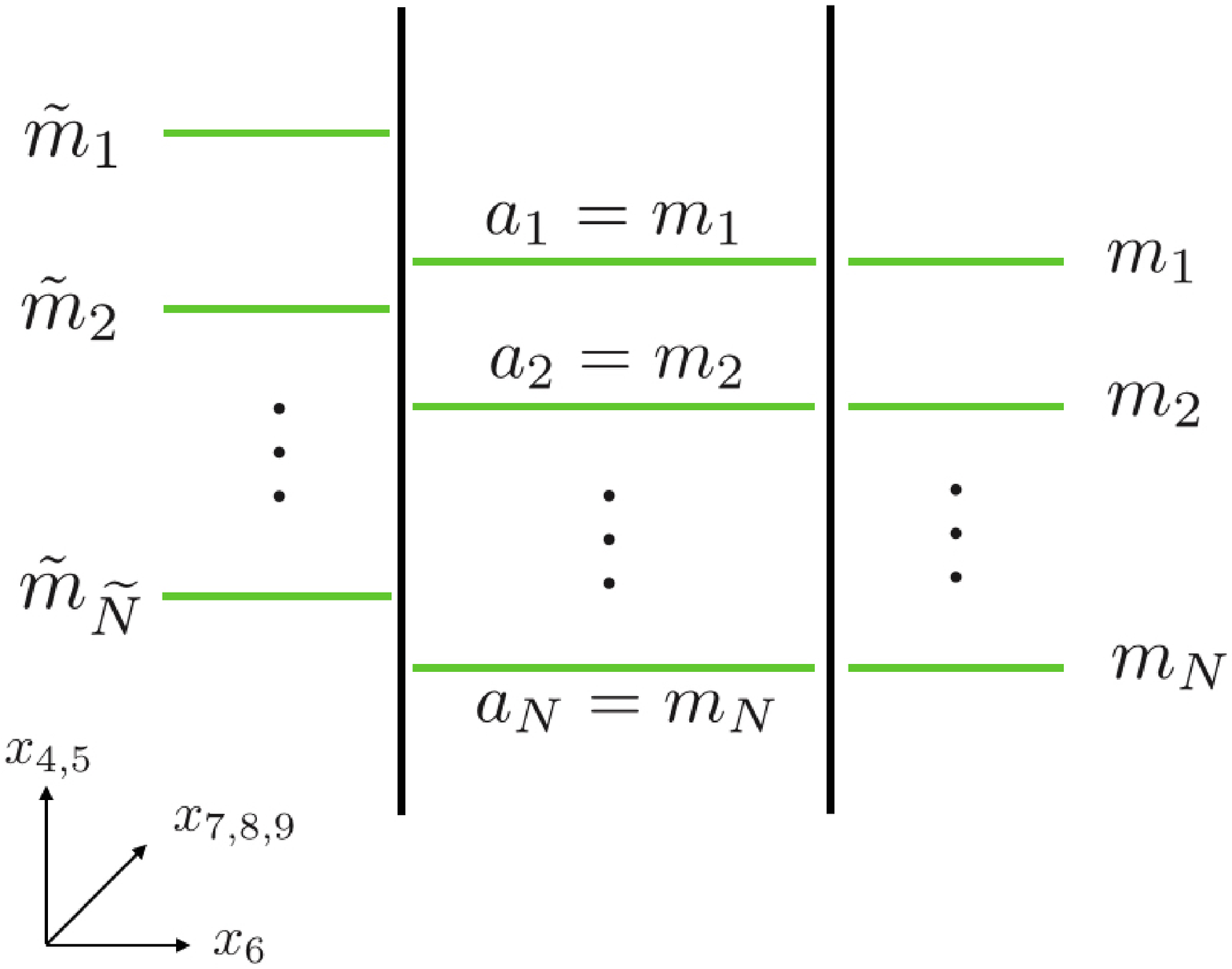}
\caption{The root of the Higgs branch}
\label{fig:rhb}
\end{minipage}
\end{figure}

The 4d theory we consider is $\mathcal{N}=2$ $U(N)$ gauge theory 
with $N$ fundamental and $\tilde{N}$ anti-fundamental hypermultiplets,
which can be realized as the D4 branes worldvolume theory 
in the Hanany--Witten brane setup (Fig.\,\ref{fig:HWbrane}).
The root of the Higgs branch can be realized 
by connecting the D4 branes to the outer flavor branes (Fig.\,\ref{fig:rhb}).
This configuration corresponds to the point on the vacuum moduli space 
at which the Coulomb branch parameters $\vec{a}$ coincide 
with the masses of the fundamental hypermultiplets:
\beq
\vec{a} = \vec{m}. 
\label{eq:RHB}
\eeq

\begin{figure}
\begin{minipage}{0.5\hsize}
\centering
\centering
\includegraphics[width=7cm]{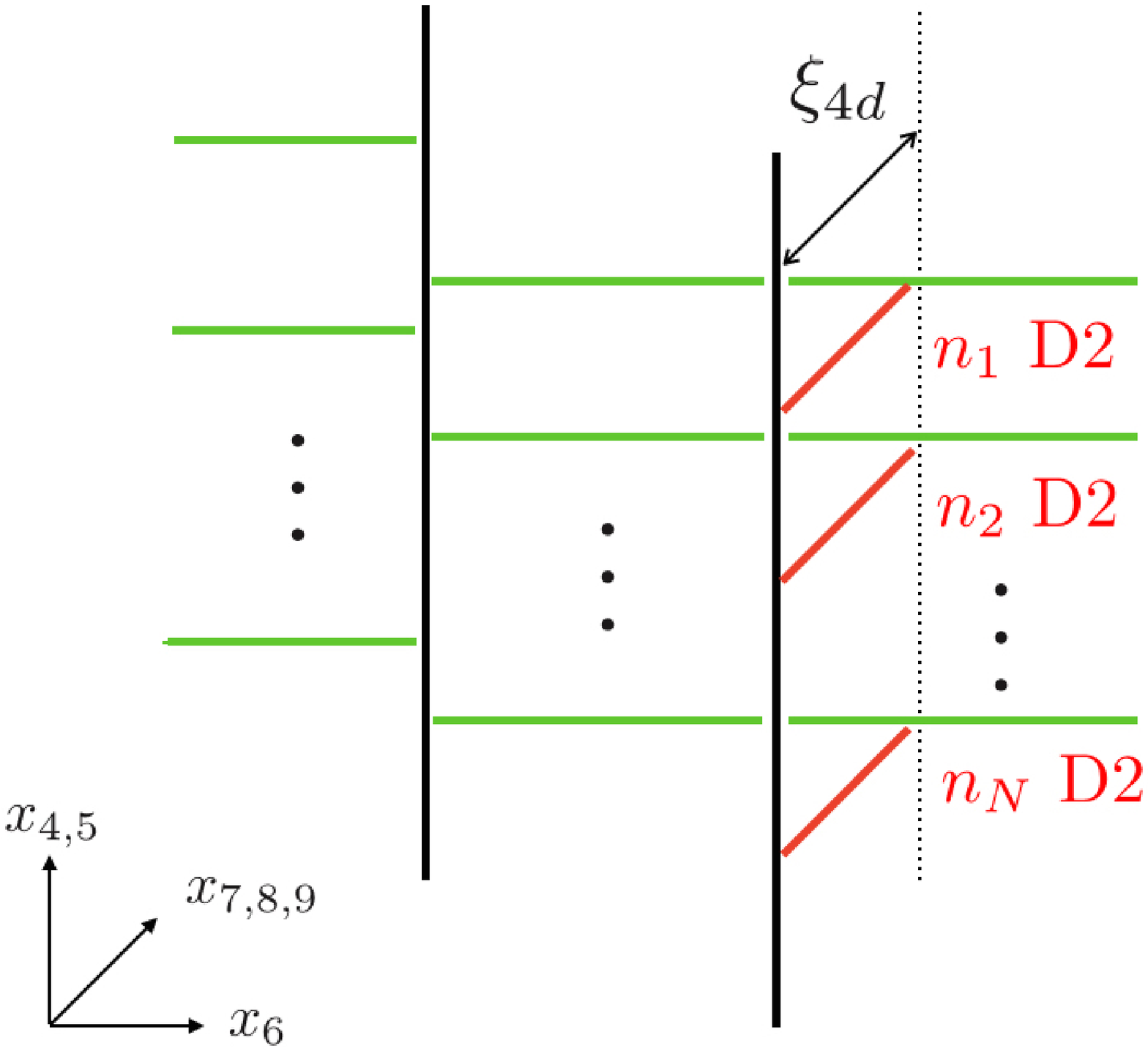}
\caption{Vortex strings in the Higgs phase.}
\label{fig:three}
\end{minipage}
\begin{minipage}{0.5\hsize}
\centering
\centering
\includegraphics[width=7cm]{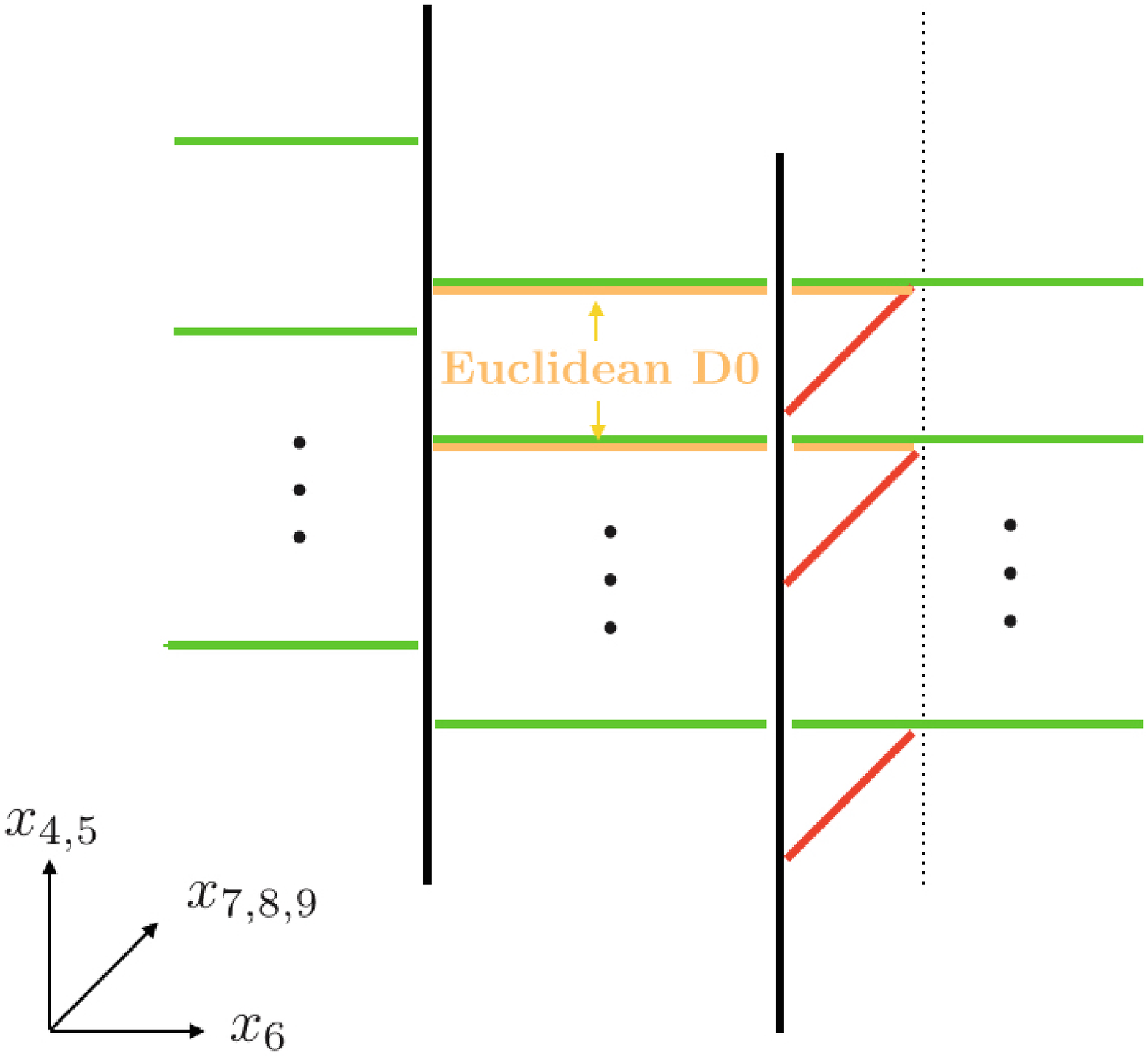}
\caption{Instantons on vortex strings.}
\label{fig:inst}
\end{minipage}
\end{figure}

The FI parameter $\xi_{4d}$ can be introduced by shifting 
one of the NS5 branes in the $x^{7,8,9}$ direction. 
Then the theory is forced onto the Higgs branch, 
where there exist half BPS vortex strings corresponding to 
D2 branes stretched between the NS5 and D4 branes (Fig.\,\ref{fig:three}). 
The worldsheet effective dynamics of $n$ vortex strings is described by 
2d $\mathcal{N}=(2,2)$ $U(n)$ gauge theory with one adjoint, 
$N$ fundamental and $\tilde{N}$ anti-fundamental chiral multiplets. 

In the M-theory picture \cite{Witten:1997sc}, the D4 and NS5 branes 
at a generic point on the Coulomb branch are identified with 
a single M5 brane described by the Seiberg--Witten curve. 
On the other hand, 
the M5 brane splits into two sheets at the root of Higgs branch \cite{Dorey:1999zk},  
so that the Seiberg--Witten curve $F(v,t)$ takes the following factorized form: 
\beq
F(v,t) = P(v,t) Q(t),
\label{eq:SW}
\eeq
where the polynomials $P(v,t)$ and $Q(t)$ are given by
\beq
P(v,t) \equiv \prod_{a=1}^{N} (v - m_a) - t  \prod_{\tilde{a}=1}^{\tilde{N}} ( v - \tilde{m}_{\tilde{a}} ) , \hs{10}
Q(t) \equiv t - \Lambda^{N-\tilde{N}}.
\eeq
One can regard $P(v,t)=0$ and $Q(t)=0$ as curves which 
determine how the two M5 branes are embedded in $\R^3 \times S^1$
parameterized by $x_4 + i x_5 \propto v$ and 
$x_6 + i x_{10} \propto \log (t/\Lambda^{N-\tilde{N}})$.

The D2 branes, which corresponds to the BPS vortex strings, 
are lifted to M2 branes stretched between the M5 branes. 
Their positions $x_4 + i x_5 \propto - v_i ~(i=1,\cdots,n)$ can be determined from $P(v,t)=Q(t)=0$, that is, 
\beq
P(v_i, t = \Lambda^{N-\tilde{N}}) = 
\prod_{a=1}^{N} (v_i - m_a) - \Lambda^{N-\tilde{N}} \prod_{\tilde{a}=1}^{\tilde{N}} (v_i - \tilde{m}_{\tilde{a}}) = 0.
\label{eq:M2-brane}
\eeq
In the 2d vortex worldsheet theory, this equation appears as the effective twisted $F$-term equation 
which determines the eigenvalues of the adjoint scalar $\sigma_i \equiv - v_i$ in the 2d vector multiplet. 

It has been known that there is a correspondence between 
the BPS objects in two and four dimensions.
At the root of the Higgs branch in four dimensions, 
there exist BPS monopoles (and dyons), 
which are confined and trapped by vortex strings.  
They are identified with BPS (dyonic) domain walls 
in the effective vortex worldsheet theory \cite{Tong:2003pz,Nitta:2010nd}. 
Furthermore, we can show an exact agreement of 
the 2d/4d SUSY central charges for the BPS objects
as a consequence of the close relationship between 
the Seiberg--Witten curve Eq.\eqref{eq:SW} and the twisted $F$-term equations \eqref{eq:M2-brane}. 
This agreement implies that the non-perturbative corrections from 
instantons (Euclidean D0-branes in Fig.\,\ref{fig:inst})
are exactly identical in four and two dimensions. 

In this section, we show the following equivalence between 
the 2d vortex partition function and the 4d instanton partition function: 
\beq
Z_{2d,\vec{n}}(m_a, \tilde{m}_{\tilde a},\epsilon_1,\epsilon_2,\tau ) = 
A(\tau) \, Z_{4d}(a_i,m_a,\tilde{m}_{\tilde a},\epsilon_1,\epsilon_2,\tau) \Big|_{\vec a = \vec m - \epsilon_1 \vec{n}} ,
\label{eq:2d_4d}
\eeq
where $\vec{n}=(n_1,\cdots,n_N) \in \Z_{\geq 0}^N$ is a vector which specifies a vacuum in 2d theory
and $U(1)^N \subset U(N)$ magnetic fluxes of the vortex strings in 4d theory (see below for details). 
$A(\tau)$ is a function which is independent of $\vec{n}$,
so that it does not appear in the relation 
between the ratios of the partition functions for different $\vec{n}$
\beq
\frac{Z_{2d,\vec{n}}}{Z_{2d,\vec{n}'}} = \frac{Z_{4d, \,\vec a = \vec m - \epsilon_1 \vec{n}}}{Z_{4d, \,\vec a = \vec m - \epsilon_1 \vec{n}'}}. 
\eeq
As a consequence of this relation, we can prove several known 2d/4d relations. 
When $\epsilon_2$, the $\Omega$-deformation parameter along the vortex worldsheet, 
is turned off, the above relation reduces to 
\beq
\widetilde{\mathcal W}_{\vec{n}} - \widetilde{\mathcal W}_{\vec{n}'} = \lim_{\epsilon_2 \rightarrow 0} \left( - \frac{\epsilon_2}{2\pi} \log \frac{Z_{4d,\,\vec a = \vec m - \epsilon_1 \vec{n}}}{Z_{4d,\,\vec a = \vec m - \epsilon_1 \vec{n}'}} \right), 
\label{eq:2d_4d_1}
\eeq
where we have used the relation between the vortex partition function and the on-shell value of the superpotential
\beq
Z_{2d, \vec{n}} = \exp \left( - \frac{2\pi}{\epsilon_2} \widetilde{\mathcal{W}}_{\vec{n}} + \mathcal{O}(1) \right).
\eeq
Eq.\,\eqref{eq:2d_4d_1} indicates the correspondence between 
the 2d domain wall central charge and the Nekrasov--Shatashvili limit of the instanton partition function \cite{Dorey:2011pa, Chen:2011sj}.
Furthermore, if both of the $\Omega$-deformation parameters $(\epsilon_1, \epsilon_2)$ are turned off, 
the relation reduces to that for the SUSY central charges 
for the 2d domain walls between the vacua $\vec{n}$ and $\vec{n}'$ 
and the 4d monopoles with magnetic charge $\vec{n}-\vec{n}'$ 
\cite{Dorey:1998yh, Dorey:1999zk, Hanany:2004ea, Shifman:2004dr}:
\beq
\lim_{\epsilon_1 \rightarrow 0} \left( \widetilde{\mathcal W}_{\vec{n}} - \widetilde{\mathcal W}_{\vec{n}'} \right) ~=~ - \frac{1}{2\pi} (\vec{n} - \vec{n}') \cdot \frac{\p}{\p \vec a} \mathcal{F},
\eeq
where we have used the following relation 
between the Nekrasov partition function 
and the 4d prepotential $\mathcal{F}$ 
\beq
Z_{4d} = \exp \left( \frac{1}{\epsilon_1 \epsilon_2} \mathcal{F} + \mathcal O(1) \right).
\eeq
Therefore, the equivalence of the partition functions Eq.\,\eqref{eq:2d_4d} is 
one of the most fundamental relations in this type of 2d/4d correspondence. 
In the following, we show Eq.\,\eqref{eq:2d_4d} by using the differential equation discussed in the previous section. 

\subsection{The Vortex Worldsheet Theory: Hanany--Tong Model}
Let us first discuss the 2d side of the correspondence, 
i.e. the effective wolrdsheet theory of $n$ vortex strings. 
This model, called the Hanany--Tong model \cite{Hanany:2003hp}, is described by 
2d $\mathcal{N}=(2,2)$ $U(n)$ gauge theory coupled to one adjoint, 
$N$ fundamental and $\tilde{N}$ anti-fundamental chiral multiplets. 
Their scalar components are denoted by $B$ ($n$-by-$n$ matrix), 
$I$ ($n$-by-$N$ matrix) and $J$ ($\tilde{N}$-by-$n$ matrix), respectively. 
The gauge coupling constant in four dimensions is identified with 
the FI parameter in two dimensions and they are combined with the $\theta$-angles 
\beq
\tau ~=~ \frac{\theta_{\rm 2d}}{2\pi} + i r_{\rm 2d} ~=~ \frac{\theta_{\rm 4d}}{2\pi} + i \frac{4 \pi}{g_{4d}^2}.
\eeq 
One of the 4d $\Omega$-deformation parameters $\epsilon_1$ corresponds to
the twisted mass for the adjoint chiral multiplet and 
the other parameter $\epsilon_2$ plays the role of $\Omega$-deformation parameter in two dimensions. 

The conditions for the supersymmetric vacua are given by
\beq
&[B,B^\dagger] + I I^\dagger - J^\dagger J = \displaystyle \frac{4 \pi}{g_{4d}^2}. & \\
&[\sigma, B] = \epsilon_1 B, \hs{10} \sigma I + I M = 0 , \hs{10} J \sigma + (\tilde{M} - \epsilon_2) J = 0, &
\eeq
where $M={\rm diag}(m_1,\cdots,m_N)$ and 
$\tilde{M} - \epsilon_2 ={\rm diag}(\tilde{m}_1 - \epsilon_2,\cdots,\tilde{m}_{\tilde{N}} - \epsilon_2)$ are
the twisted mass matrices for the chiral multiplets and the anti-chiral multiplets, respectively. 
The classical vacua of this model correspond to 
the BPS configurations of $n$ vortex strings in the 4d $\Omega$-background. 
For the Abelian vortices ($N=1$), there exist unique vacuum, 
in which the scalar expectation values are given by 
\renewcommand\arraystretch{0.8}
\beq
{\arraycolsep = 1.2mm
B \propto  
\ba{cccc} 
0 & 1 & & \\
& \ddots & \ddots \\
& & 0 & 1 \\
& & & 0 
\ea, 
\hs{10}
I \propto \ba{c} 0 \\ \vdots \\ 0 \\ 1 \ea},
\eeq
and $J=0$. For general $N$, vacuum configurations can be obtained 
by decomposing the matrices into $N$ blocks and embedding the above solution. 
The sizes $n_a~(a=1,\cdots,N)$ of the diagonal blocks of $B$ are arbitrary as long as the total size is $n$.  
Thus, the vacua are labeled by a set of non-negative integers 
$\vec{n}=(n_1,\cdots,n_N)$ such that $\sum_{a=1}^N n_a = n$. 

The saddle point configurations in this model are 
BPS vortex configurations corresponding to 4d Yang--Mills instantons trapped on the vortex worldsheets. 
In each vacuum, the saddle point condition Eq.\,\eqref{eq:saddle_cond} 
for the classical eigenvalues of the vector multiplets scalar $\sigma$ 
takes the form 
\beq
\sigma_{a}^{q} - \sigma_{a}^{q+1} = \epsilon_1 - k_{a}^{q} \epsilon_2, \hs{5} 
\sigma_{a}^{n_a} = - m_a - k_a^{n_a} \epsilon_2, \hs{10} (q=1,\cdots, n_a,~ a=1,\cdots,N), 
\eeq
where we have decomposed the label of the eigenvalues $\sigma_i\,(i=1,\cdots,n)$ as $\sigma_{a}^q$. 
Solving this condition with respect to $\sigma_{a}^q$, we find that 
\beq
\sigma_a^q ~=~ - m_a + (n_a-q) \epsilon_1 - \lambda_{a}^q \epsilon_2, \hs{10} 
(\lambda_{a}^q - \lambda_a^{q+1} = k_a^{q},~ \lambda_{a}^{n_a+1}=0).
\label{eq:sigmaHTvortex}
\eeq
Since the winding numbers $k_{a}^q$ are all non-negative,  
the integers $\lambda_{a}^q$ satisfy 
\beq
\lambda_{a}^1 \geq \lambda_{a}^2 \geq \cdots \geq  \lambda_{a}^{n_a} \geq \lambda_{a}^{n_a+1}=0.
\eeq 
In other words, the saddle point vortex configurations in the vacuum $\vec{n}$ 
are labeled by a set of $N$ Young tableaux $Y_a=(\lambda_{a}^1, \lambda_{a}^2, \cdots)$ 
which have at most $n_a$ rows (since $\lambda_{a}^q=0$ for $q \geq n_a + 1$).

The general formula Eq.\,\eqref{eq:integral_NA} implies that 
the vortex partition function in the vacuum $\vec{n}$ is given by
the following integral along a path enclosing all the poles at Eq.\,\eqref{eq:sigmaHTvortex}:
\beq
Z_{2d, \vec{n}}(\tau_1,\cdots,\tau_n) = \int_{C_{\vec{n}}} \prod_{i=1}^n \left[ \frac{d \sigma_i}{2 \pi i \epsilon_2} 
\exp \left(-\frac{2\pi i \sigma_i \tau_i}{\epsilon_2} \right) \right] Z_{g} Z_{ad} Z_f Z_{af}, 
\label{eq:HT_int}
\eeq
where $Z_{g}$, $Z_{ad}$, $Z_{f}$ and $Z_{af}$ are the contributions 
from the gauge, adjoint, fundamental and anti-fundamental degrees of freedom
\begin{align}
 Z_{g} &~=
 \prod_{1 \le i \neq j \le n}
 \Gamma \left(\frac{\sigma_i - \sigma_j}{\epsilon_2}\right)^{-1}, 
&Z_{ad} &~=~ \prod_{i=1}^n \prod_{j=1}^n \Gamma\left( \frac{\sigma_i - \sigma_j - \epsilon_1}{\epsilon_2}\right), \\
Z_{f} &~=~ \prod_{i=1}^n \prod_{a=1}^N ~~ \Gamma \left( \frac{\sigma_{i} + m_a}{\epsilon_2} \right), \hs{-7}
&Z_{af} &~=~ \prod_{i=1}^{n} \prod_{\tilde{a}=1}^{\tilde N} \Gamma \left( - \frac{\sigma_{i} + \tilde{m}_{\tilde{a}}-\epsilon_2}{\epsilon_2} \right).
\end{align}
Since there is no fractional vortex in this model, 
the vortex partition function simply factorizes into perturbative and vortex parts
\beq
Z_{2d, \vec{n}} = Z_{2d, \vec{n}}^{pert} \, Z_{2d, \vec{n}}^{vortex}.
\eeq
The perturbative part is given by
the residue of the integrand in Eq.\,\eqref{eq:HT_int} 
at the pole corresponding to the vacuum configuration
\beq
Z_{2d,\vec{n}}^{pert} = \prod_{a=1}^N \prod_{q=1}^{n_a} \Bigg[ \exp \left(- \frac{2\pi i \tau_a^q \underline{\sigma}_{a}^q}{\epsilon_2} \right)\prod_{b=1}^N \Gamma \left( \frac{\underline{\sigma}_{a}^q + m_b - n_{b} \epsilon_1}{\epsilon_2} \right) \prod_{\tilde a=1}^{\widetilde N} 
\Gamma \left( -\frac{\underline{\sigma}_{a}^{q} + \tilde{m}_{\tilde a} -\epsilon_2}{\epsilon_2} \right) \Bigg], 
\label{eq:2d_pert}
\eeq
where $\underline{\sigma}_{a}^{q}$ are the eigenvalues in the classical vacuum configuration
\beq
\underline{\sigma}_{a}^{q}= - m_a - (q-n_a) \epsilon_1.
\label{eq:vev_sigma}
\eeq
On the other hand, the vortex part is the collection of the contributions 
from all the saddle points, 
which can be written as a sum over the Young tableaux 
\beq
Z_{2d,\vec{n}}^{vortex}(\tau_1,\cdots,\tau_n) ~=~ 
\sum_{\vec{Y}} \exp \bigg( \sum_{a=1}^N \sum_{q=1}^{n_a} 2 \pi i \lambda_a^q \tau_a^q \bigg) Z_{2d,\vec{n}}^{\vec{Y}}.
\label{eq:inst_expansion}
\eeq 
In order to compare this quantity with the Nekrasov partition function in four dimensions, 
it is convenient to use the differential equations discussed in the previous section
\beq
P_{i} (\hat{\sigma}_1,\cdots,\hat{\sigma}_n ) \, Z_{2d,\vec{n}}(\tau_1,\cdots,\tau_n) ~=~ 0, \hs{10} (i=1,\cdots,n), 
\label{eq:HT_diffeq}
\eeq
where $\hat{\sigma}_i$ are the differential operator
\beq
\hat{\sigma}_i \equiv - \frac{\epsilon_2}{2\pi i} \frac{\p}{\p \tau_i}.
\eeq
The explicit forms of the polynomials $P_i(\hat{\sigma}_i)$ can be obtained 
from the invariance of the integral Eq.\,\eqref{eq:HT_int} 
under the shift of the integration path $\sigma_i \rightarrow \sigma_i + 1$:
\beq
P_i &=& Q^+_{i} - \epsilon_2^{N-\tilde N} \exp \left( 2 \pi i \tau_i \right) Q^-_{i}, 
\eeq
where $Q_i^{\pm}~(i=1,\cdots,n)$ are the following polynomials
\beq
Q_{i}^{+}(\hat{\sigma}_1,\cdots,\hat{\sigma}_n) &=& 
\left[ \prod_{j \not = i} \left( \hat \sigma_{i} - \hat \sigma_{j} - \epsilon_1 \right) \left( \hat \sigma_{i} - \hat \sigma_{j} + \epsilon_2 \right) \right] 
\left[ \prod_{a=1}^N ( \hat \sigma_i + m_a ) \right] , \\
Q_{i}^{-}(\hat \sigma_1,\cdots,\hat \sigma_n) &=& 
\left[ \prod_{j \not = i} \left( \hat \sigma_{i} - \hat \sigma_{j} + \epsilon_1 \right) \left( \hat \sigma_{i} - \hat \sigma_{j} - \epsilon_2 \right) \right]
\left[ \prod_{\tilde{a}=1}^{\tilde N} ( \epsilon_2 - \hat \sigma_i - \tilde{m}_{\tilde{a}} ) \right]  .
\eeq
Once the perturbative part Eq.\,\eqref{eq:2d_pert} is given, 
the differential equations can be rewritten in terms of the vortex part:
\beq
P_i(\hat{\sigma}_i - \underline{\sigma}_{\, i}) \, Z_{2d,\vec{n}}^{vortex} = 0,
\label{eq:diff_vortex}
\eeq
where the constants $\underline{\sigma}_{\,i}$ denote the classical VEVs Eq.\,\eqref{eq:vev_sigma}. 
This differential equation completely determine the vortex part $Z_{2d,\vec{n}}^{vortex}$
since all the coefficients $Z_{2d,\vec{n}}^{\vec{Y}}$ in the expansion Eq.\,\eqref{eq:inst_expansion}
can be uniquely determined by setting the first term $Z_{2d,\vec{n}}^{\vec{Y} = \emptyset} = 1$ and 
recursively solving Eq.\,\eqref{eq:diff_vortex} order by order.
We will use this fact to prove the equivalence 
between the Nekrasov partition function and the vortex partition function.

Let us remark that in the limit $\epsilon_2 \rightarrow 0$, 
the differential equations Eq.\,\eqref{eq:HT_diffeq} for $\tilde N =N$ reduces to 
\beq
\prod_{j \not = i} \frac{\left( \sigma_{i} - \sigma_{j} - \epsilon_1 \right) }{\left( \sigma_{i} - \sigma_{j} + \epsilon_1 \right)} =
(-1)^{N} e^{2 \pi i \tau} \frac{\prod_{\tilde{a}=1}^{N} ( \sigma_i + \tilde{m}_{\tilde{a}})}{\prod_{a=1}^N ( \sigma_i + m_a )} .
\eeq
This is the twisted $F$-term equation of the Hanany--Tong model 
and as pointed out in \cite{Dorey:2011pa, Chen:2011sj}, 
this agrees with the Bethe ansatz equation of the quantum $SL(2,\R)$
spin chain, in which $\epsilon_1$ plays a role of the Planck constant.
See also~\cite{Nekrasov:2009rc,Nekrasov:2012xe,Nekrasov:2013xda}.
On the other hand, as we have seen in the previous section, 
$\epsilon_2$ is another Planck constant characterizing 
the commutation relation Eq.\,\eqref{eq:quantization}.
In this sense, the system of the differential equations \eqref{eq:HT_diffeq} 
has two Planck constants associated with the two different quantization conditions. 
If both the Planck constants are turned off, 
the differential equations \eqref{eq:HT_diffeq} reduce to 
the twisted $F$-term equations \eqref{eq:M2-brane} which describe the M2-branes. 
In this sense, the differential equations \eqref{eq:HT_diffeq} can be viewed 
as a ``doubly quantized'' version of the twisted $F$-term equations. 

\subsection{Comparison with Nekrasov Partition Function} 
\paragraph{Perturbative part \\}
Let us first compare the perturbative part of the vortex partition function Eq.\,\eqref{eq:2d_pert}
and that of the Nekrasov partition function
\begin{align}
 Z_{4d}^{pert}
 & =
 \exp \left( -\frac{\vec a \cdot \vec a}{\epsilon_1 \epsilon_2} \pi i \tau_{0} \right) 
 \prod_{1 \le a \neq b \le N}
 \Gamma_2(a_b - a_a \mid \epsilon_1, \epsilon_2)^{-1}
 \nonumber \\
 & \qquad \times
 \prod_{b=1}^N
 \left[
 \prod_{a=1}^N
  \Gamma_2 ( a_{b} - m_a \mid \epsilon_1,\epsilon_2)
 \prod_{\tilde{a}=1}^{\tilde{N}}
  \Gamma_2 ( a_b + \epsilon_+ - \tilde{m}_{\tilde{a}} \mid
   \epsilon_1,\epsilon_2)
 \right].
\label{eq:4d_pert} 
\end{align}
where $\epsilon_+ \equiv \epsilon_1 + \epsilon_2$ and $\Gamma_2(x|\epsilon_1,\epsilon_2)$ is the double gamma function
\beq
\Gamma_2(x|\epsilon_1,\epsilon_2) = \prod_{n=0}^\infty \prod_{m=0}^\infty \frac{\Lambda_0}{x + n \epsilon_1 + m \epsilon_2}.
\eeq
This factor is obtained using the double zeta function in the
same way as \eqref{eq:zeta_ref}.

Now let us focus on the root of Higgs branch. 
In the presence of vortex strings in the $\Omega$-background, 
the Coulomb branch parameters $\vec{a}$ are shifted
from those in the undeformed theory Eq.\,\eqref{eq:RHB} as\footnote{
A similar phenomenon can also be seen for the vortices in the 2d $\Omega$-background discussed in Sec.\,\ref{sec:part_func}:
\beq
\sigma \big|_{z=0} ~=~ - m_a - k \epsilon. \notag
\eeq
}
\beq
\vec{a} = \vec{m} - \epsilon_1 \vec{n}.
\label{eq:RHB_Omega}
\eeq
Plugging this relation into Eq.\,\eqref{eq:4d_pert}, 
we can show that the classical part, the vector-fundamental part, and the anti-fundamental part
are respectively given by
\beq
\exp \Big( -\frac{\vec a \cdot \vec a}{\epsilon_1 \epsilon_2} \pi i \tau_{0} \Big) \ &=& 
\exp \bigg[ - \frac{2 \pi i \tau_0}{\epsilon_2} \sum_{a=1}^N \sum_{q=1}^{n_a} \underline{\sigma}_a^q \bigg] e^{-\frac{\pi i \tau_{0}}{\epsilon_1 \epsilon_2} ( \vec m \cdot \vec m - n \epsilon_1^2 )},   \\
\frac{\Gamma_2 ( a_{b} - m_a |\epsilon_1,\epsilon_2) }{\Gamma_2(a_{b}-a_{a}|\epsilon_1,\epsilon_2)} ~ &=& \ \prod_{q=1}^{n_a} \Gamma_1 \left( \underline{\sigma}_a^q + m_{b} - n_b \epsilon_1 | \epsilon_2 \right), \label{eq:vec-fund_cancel}\\
\Gamma_2 ( a_b + \epsilon_+ - \tilde{m}_{\tilde{a}} | \epsilon_1,\epsilon_2) \hs{-1} &=& 
\bigg[ \prod_{q=1}^{n_b} \Gamma_1 ( \epsilon_2 - \underline{\sigma}_b^q - \tilde{m}_{\tilde{a}} |\epsilon_2) \bigg]
\Gamma_2 ( \epsilon_+ + m_b - \tilde{m}_{\tilde{a}} | \epsilon_1,\epsilon_2),
\eeq
where we have used the relations between the Gamma functions
\beq
\Gamma_2(x|\epsilon_1,\epsilon_2) =  \Gamma_1(x|\epsilon_2) \Gamma_2(x+\epsilon_1|\epsilon_1,\epsilon_2), \hs{5}
 \Gamma_1(x|\epsilon_2) \equiv \frac{1}{\sqrt{2 \pi}}  \left( \frac{\Lambda_0}{\epsilon_2} \right)^{-\frac{x\,}{\epsilon_2}+\frac{1}{2}} \Gamma \left( \frac{x}{\epsilon_2} \right).
\eeq 
For the cancellation of the vector-fundamental part
\eqref{eq:vec-fund_cancel}, in particular, for $a=b$, the diagonal
factors of the vector multiplet contribution are needed in the
expression \eqref{eq:4d_pert}.
Since such factors are independent of the Coulomb moduli and mass
parameters, we put them as a vacuum-independent factor.

Combining these three factors, 
we find that the perturbative part in four dimensions agrees 
with that in two dimensions Eq.\,\eqref{eq:2d_pert} up to an $\vec{n}$ independent factor:  
\beq
Z_{2d,\vec{n}}^{pert} (\tau_a^p = \tau) ~=~ A(\tau) Z_{4d}^{pert}(\tau) \Big|_{\vec{a} = \vec{m}-\epsilon_1\vec{n}} .
\label{eq:perturb_eq}
\eeq
where the renormalized $\tau$-parameter is given by
\beq 
\tau = \tau_0 + \frac{N - \tilde{N}}{2\pi i} \log \frac{\Lambda_0}{\epsilon_2}.
\eeq

\paragraph{Instanton part \\} 
At the root of the Higgs branch $\vec{a} = \vec{m}-\epsilon_1 \vec{n}$, 
the instanton part $Z_{4d}^{inst}(\tau)$ can be promoted to the $n$-variable function defined by
\beq
Z_{4d}^{inst}(\tau_1,\cdots,\tau_n) ~\equiv~ 
\sum_{\vec{Y}} \exp \left( \sum_{a=1}^N \sum_{q=1}^{n_a} 2 \pi i \lambda_a^q \tau_a^q \right) Z_{4d}^{\vec{Y}}. 
\eeq
We show that $Z_{4d}^{inst}(\tau_1,\cdots,\tau_n)$ satisfies the differential equation \eqref{eq:diff_vortex} 
which gives the vortex part of the 2d partition function. 

The contribution from each instanton configuration is given by 
the following integral formula
\beq
Z_{4d}^{\vec{Y}} \hs{-2} &=& \hs{-2} \int ~ \prod_{\alpha=1}^{|\vec{Y}|} \Bigg[ \frac{d \Phi_\alpha}{2 \pi i} 
\frac{\epsilon_1+\epsilon_2}{\epsilon_1\epsilon_2} 
\prod_{\beta < \alpha} \frac{(\Phi_{\alpha} - \Phi_{\beta})^2 \big\{ (\Phi_{\alpha} - \Phi_{\beta})^2 - (\epsilon_1+\epsilon_2)^2 \big\}}
{\big\{(\Phi_{\alpha} - \Phi_{\beta})^2 - \epsilon_1^2 \big\} \big\{(\Phi_{\alpha} - \Phi_{\beta})^2-\epsilon_2^2 \big\}}  \notag \\
&\times& \hs{5} 
\prod_{b=1}^N \frac{1}{(\Phi_{\alpha} + a_b)(\epsilon_1+\epsilon_2 - a_b - \Phi_\alpha)} 
\prod_{a=1}^N ( - m_a - \Phi_\alpha) \prod_{\tilde{a}=1}^{\tilde{N}} (\epsilon_1+\epsilon_2 -\tilde{m}_{\tilde{a}} - \Phi_\alpha) \Bigg],
\eeq
where the integration contours for $\Phi_{\alpha}~(\alpha=1,\cdots,|\vec{Y}|)$ 
are those which enclose the following poles 
specified by the Young tableaux $\vec{Y}$: 
\beq
\Phi_{b}^{(p,q)} ~=~ - a_b - (q-1) \epsilon_1 - (p -1) \epsilon_2,
\label{eq:pole}
\eeq
where we have relabeled $\Phi_{\alpha}$ as 
$\Phi_{b}^{(p,q)}~(b=1,\cdots,N, ~(p,q) \in Y_b)$. 

\begin{figure}[h]
\centering
\includegraphics[width=80mm]{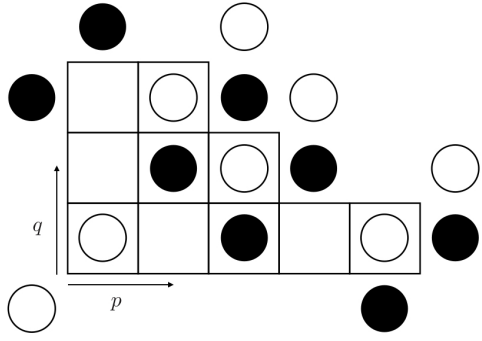}
\caption{The exponents $I_b{(p,q)}$ for a tableau: $I_b{(p,q)} =1$ and $I_b{(p,q)}=-1$ at the points with the white and black circles, respectively.}
\label{fig:tableau}
\end{figure}

The differential equation for $Z_{4d}^{inst}(\tau_1,\cdots,\tau_n)$ can be obtained 
from a recursion relation for $Z_{4d}^{\vec{Y}}$, which can be derived in the following way. 
Let $\vec{Y}'$ be the Young tableaux obtained by adding a box to $\vec{Y}$. 
The integral representation for $Z_{4d}^{\vec{Y}'}$ has one more integration variable than that for $Z_{4d}^{\vec{Y}}$. 
The contour for the new integration variable $\Phi$ should be 
chosen so that it encloses the pole corresponding to the added box, 
while for the other variable, 
the contours are kept the same as those for $Z_{4d}^{\vec{Y}}$.
Thus, $Z_{4d}^{\vec{Y}}$ is contained in $Z_{4d}^{\vec{Y}'}$ as its $\Phi$-independent part 
and their ratio takes the form 
\beq
\frac{Z_{4d}^{\vec Y'}}{Z_{4d}^{\vec Y}} &=& \int \frac{d \Phi}{2 \pi i} \frac{\epsilon_1+\epsilon_2}{\epsilon_1\epsilon_2} 
\Bigg[ \prod_{b=1}^N  \prod_{p,q = 0}^{\infty} \Big\{ \Phi + a_b + (q-1) \epsilon_1 + (p -1) \epsilon_2 \Big\}^{I_b{(p,q)}}  \notag \\
&\times&
\prod_{b=1}^N \frac{1}{(\Phi + a_b)(\epsilon_1 + \epsilon_2 - a_b - \Phi)} 
\prod_{a=1}^N (- m_a - \Phi) \prod_{\tilde{a}=1}^{\tilde{N}} (\epsilon_1+\epsilon_2 -\tilde{m}_{\tilde{a}} - \Phi) \Bigg].
\label{eq:ratio0}
\eeq
The exponents $I_b{(p,q)}$ can be determined from the Young tableau $Y_b$ as (see Fig. \ref{fig:tableau})
\beq
I_b{(p,q)} = \left[ 2 \rho_b^{(p,q)} + \rho_b^{(p+1,q+1)} + \rho_b^{(p-1,q-1)} \right] 
- \left[ \rho_b^{(p-1,q)} + \rho_b^{(p+1,q)} + \rho_b^{(p,q+1)} + \rho_b^{(p,q-1)} \right],
\eeq
where $\rho_b^{(p,q)}$ is one or zero depending on 
if $(p,q) \in Y_b$ or $(p,q) \not \in Y_b$.  
Eq.\,\eqref{eq:ratio0} can be further simplified by using the relation
\beq
\prod_{p=0}^\infty \Big\{ \Phi + a_b + (q-1) \epsilon_1 + (p-1) \epsilon_2 \Big\}^{I_b{(p,q)}} = 
\frac{
(\Phi-\Phi_{b}^{q})
(\Phi - \Phi_{b}^{q-1} + \epsilon_1+\epsilon_2)}
{(\Phi - \Phi_{b}^{q+1} - \epsilon_1) 
(\Phi - \Phi_{b}^{q} + \epsilon_2)}, 
\label{eq:inst_integrand}
\eeq
where $\Phi_{b}^q$ is given by 
\beq
\Phi_b^q = - a_b - (q-1) \epsilon_1 - (\lambda_b^q-1) \epsilon_2, \hs{10} \left( \mbox{$\lambda_b^{q} = 0$ for $q = 0,\, -1$} \right).
\eeq

Now let us focus on the root of the Higgs branch. 
When $a_b = m_b - \epsilon_1 n_b$, the pole corresponding to the box at $(1,n_b+1)$ disappears. 
Thus, all the contributions from $Y_b$ with more than $n_b$ rows vanish 
and the infinite product over $q$ in Eq.\,\eqref{eq:ratio0} terminates 
at $q = n_b+1$ for $Y_b$ with $\lambda_b^{n_b+1} = 0$. 
Integrating Eq.\,\eqref{eq:ratio0} around the pole at $\Phi = \Phi_b^p - \epsilon_2$, 
we find taht
\beq
\frac{Z_{\vec Y'}}{Z_{\vec Y}} = (-1)^{\tilde{N}}
\prod_{(a,q)\neq(b,p)}
\frac{
\big( \Phi_{b}^{p} - \Phi_{a}^{q} - \epsilon_2 \big)
\big( \Phi_{b}^{p} - \Phi_{a}^{q} + \epsilon_1 \big)}
{(\Phi_{b}^{p} - \Phi_{a}^{q} - \epsilon_1 - \epsilon_2 ) 
\big( \Phi_{b}^{p} - \Phi_{a}^{q} \big)} 
\frac{\prod_{\tilde{a}=1}^{\tilde{N}} (\Phi_{b}^{p} + \tilde{m}_{\tilde{a}}-\epsilon_1-2\epsilon_2) }{\prod_{a=1}^N (\Phi_{b}^{p} + m_a - \epsilon_1 - 2\epsilon_2)}.
\label{eq:inst_recursion}
\eeq 
This is the recursion relation which can be obtained by substituting 
\beq
Z_{4d}^{inst}(\tau_1,\cdots,\tau_n) ~\equiv~ 
\sum_{\vec{Y}} \exp \left( \sum_{a=1}^N \sum_{q=1}^{n_a} 2 \pi i \lambda_a^q \tau_a^q \right) Z_{4d}^{\vec{Y}}, 
\eeq
into the differential equation Eq.\,\eqref{eq:diff_vortex}. 
Therefore, $Z_{4d}^{inst}(\tau_1,\cdots,\tau_n)$ satisfies the differential equation
which uniquely determines the vortex part of the 2d partition function, 
and hence the non-perturbative parts of the 2d and 4d partition functions are identical to each other
\beq
Z_{2d, \vec{n}}^{vortex}(\tau_1,\cdots,\tau_n) ~=~ Z_{4d}^{inst}(\tau_1,\cdots,\tau_n) \Big|_{\vec{a}=-\vec{m}-\epsilon_1 \vec{n}}.
\eeq
In conjunction with the perturbative part Eq.\,\eqref{eq:perturb_eq}, 
this relation (with $\tau_i = \tau$ for all $i$) shows 
the equivalence of the 2d/4d partition functions Eq.\,\eqref{eq:2d_4d}.

Let us remark that the differential equation for the Nekrasov partition
function at the root of Higgs branch, discussed in this paper, is a
special case of that studied in \cite{Kanno:2012hk,Kanno:2013aha}.
From this point of view, the simplification due to the root of Higgs
branch would be related to the Nekrasov--Shatashvili limit of the
corresponding Hecke algebra~\cite{Bourgine:2014tpa}.

\section{Conclusion}\label{sec:summary}

In this paper, we have considered the $\Omega$-background formulation in
two dimensions, and applied the localization scheme to obtain the exact
formula of the partition function.
We have evaluated the one-loop determinant in the two-dimensional
$\Omega$-background configuration, and obtained the combinatorial and
integral representations of the partition function. 
We have observed the resultant expression is essentially equivalent to
the exact result for the two-dimensional hemisphere.
We have then studied the differential equation which the partition
function satisfies through the quantization of the classical algebraic
geometry.
We have considered the vortex worldsheet theory as an application of the
exact formula obtained in this paper.
We have observed the agreement of the two- and four-dimensional theories
at the level of the partition function, by taking the root of Higgs branch.

Let us comment on some possibilities of future works.
The first interesting direction is generalization to the
three-dimensional theory on $\mathbb{R}^2 \times S^1$, which is the
$q$-deformation of the current situation.
For such a theory, a similar relation between three- and
five-dimensional theories is expected to
hold~\cite{Aganagic:2013tta,Aganagic:2014oia,Aganagic:2015cta}.
See also~\cite{Nieri:2013yra,Nieri:2013vba}.
If we obtain the three-dimensional partition function,
we could check the 3d/5d
correspondence in the same way as this paper.
Furthermore, to see its relation to the $q$-deformed CFT correlation
functions, we have to extend our analysis to the quiver
gauge theory in two or three dimensions.
We can start with, for example, the brane configuration considered
in~\cite{Dorey:2011pa,Chen:2011sj}, and then check the coincidence of
the partition functions more explicitly.



\subsection*{Acknowledgments} 

The authors are grateful to H.~Fuji, Y.~Matsuo, and M.~Taki for useful comments.
The work of MN is supported in part by a Grant-in-Aid for Scientific
Research on Innovative Areas ``Topological Materials Science''
(KAKENHI Grant No. 15H05855) and ``Nuclear Matter in Neutron Stars
Investigated by Experiments and Astronomical Observations'' (KAKENHI
Grant No. 15H00841) from the the Ministry of Education, Culture, Sports,
Science (MEXT) of Japan.
The work of TF and MN is also supported in part
by the Japan Society for the Promotion of Science (JSPS)
Grant-in-Aid for Scientific Research (KAKENHI Grant No. 25400268) and by
the MEXT-Supported Program for the Strategic Research Foundation
at Private Universities ``Topological Science'' (Grant No. S1511006).
The work of TK was supported in part by JSPS Grant-in-Aid for Scientific
Research (No. 13J04302) from MEXT of Japan.

\appendix

\section{Spinor Conventions}
\label{appendix:conventions}
Our convention for the anti-symmetric tensors $\epsilon^{\alpha \beta}$ and $\epsilon_{\dot \alpha \dot \beta}$ is
\beq
\epsilon^{12} = \epsilon_{\dot 1 \dot 2} = 1, \hs{10} \epsilon^{21} = \epsilon_{\dot 2 \dot 1} = -1. 
\eeq
The 4d Weyl spinor indices are raised and lowered as
\beq
\lambda^\alpha = \epsilon^{\alpha \beta} \lambda_\beta, \hs{10}
\bar{\lambda}_{\dot \alpha} = \epsilon_{\dot \alpha \dot \beta} \bar{\lambda}^{\dot \beta}.
\eeq
The spinor bilinears without spinor indices are given by
\beq
\lambda \psi = \lambda^\alpha \psi_\alpha, \hs{5} 
\bar{\lambda} \bar{\psi} = \bar{\lambda}_{\dot \alpha} \bar{\psi}^{\dot \alpha}, \hs{5} 
\bar{\lambda} \sigma_i \psi = \bar{\lambda}_{\dot \alpha} (\sigma_i)^{\dot \alpha \beta} \psi_\beta, \hs{5} 
\lambda \bar{\sigma}_i \bar{\psi} = \lambda^{\alpha}(\bar{\sigma}_i)_{\alpha \dot \beta} \bar{\psi}^{\dot \beta}.
\eeq
Let $\sigma_i = \bar{\sigma}_i~(i=1,2,3)$ be the standard Pauli matrices 
and
\beq
\sigma_4 = -\bar{\sigma}_4 = \ba{cc} i & 0 \\ 0 & i \ea.
\eeq
The rotation generators are defined by
\beq
\sigma_{ab} = -\frac{i}{4}(\bar{\sigma}_a \sigma_b - \bar{\sigma}_a \sigma_b), \hs{10}
\bar{\sigma}_{ab} = -\frac{i}{4}(\sigma_a \bar{\sigma}_b - \sigma_a \bar{\sigma}_b), 
\eeq
Under the 2d rotation $x_1 + i x_2 \rightarrow e^{i \theta} (x_1 + i x_2)$, 
2d spinors transform as
\beq
\lambda \rightarrow e^{-i \theta \sigma_{12}} \lambda, \hs{10} 
\bar{\lambda} \rightarrow e^{-i \theta \bar{\sigma}_{12}} \bar{\lambda}.
\eeq

\section{Mode Expansion around Saddle Points}\label{appendix:determinants}
Here we consider the mode expansion around the BPS background Eq\,\eqref{eq:saddle}. 
For notational simplicity, we set $g=1$ and the flavor indices are suppressed in the following. 
Let us first determine the operators $\Delta_B$ and $\Delta_F$ appearing in the linearized equations of motion
\beq
\Delta_B \Phi =0, \hs{10} \Delta_F \Psi =0. 
\eeq

The fermionic part of the action takes the form
\beq
\mathcal L_{F} = - i (\bar{\Psi} \cdot \nabla \Psi + \nabla^\ast \bar{\Psi} \cdot \Psi), \hs{10} 
\Psi = \ba{c} 
\bar{\lambda}_0 \\
\bar{\lambda}_{\bar{z}} \\
\psi_{0} \\
\psi_{\bar{z}} 
\ea, \hs{5}
\bar{\Psi} =
\ba{c} 
\lambda_0 \\
\lambda_{z} \\
\bar{\psi}_0 \\
\bar{\psi}_{z}
\ea,
\eeq 
where $\nabla$ is given by
\beq
\nabla \equiv
{\renewcommand\arraystretch{1.1}
\ba{cccc} 
\D_{\xi} & \p_{z} & \frac{\bar{\phi}}{\sqrt{2}} & 0 \\
-\p_{\bar{z}} & \D_{\bar{\xi}} & 0 & -\frac{\bar{\phi}}{\sqrt{2}} \\
\frac{\phi}{\sqrt{2}} & 0 & \D_{\bar{\xi}} & \D_z \\
0 & - \frac{\bar{\phi}}{\sqrt{2}} & - \D_{\bar{z}} & \D_{\xi}
\ea},
\eeq
and $\nabla^\ast$ is the adjoint operator of $\nabla$. 
The equations of motion for the fermionic fields is $- i \nabla \Psi =0$ and hence
\beq
\Delta_F = - i \nabla.
\eeq

The bosonic equations of motion can be written 
in terms of the supersymmetry variations of the fermionic fields:
\beq
{\renewcommand\arraystretch{1.1}
\ba{c}
-\frac{\delta S}{\delta A_{\bar \xi}} \\
\phantom{-} \frac{\delta S}{\delta A_z} \\
\frac{-i}{\sqrt{2}} \frac{\delta S}{\delta \bar{\phi}} \\
0 
\ea}
~=~
2 \nabla (\mathcal{Q} \Psi), \hs{10} 
\mathcal{Q} \Psi ~\equiv~
{\renewcommand\arraystretch{1.1}
\ba{c} 
\mathcal{Q} \bar{\lambda}_0 \\
\mathcal{Q} \bar{\lambda}_{\bar{z}} \\
\mathcal{Q} \psi_0 \\
\mathcal{Q}\psi_{\bar{z}} 
\ea}.
\label{eq:bosonic_eom}
\eeq
Note that the equation in the forth component is an identity which is automatically satisfied. 
Since $\mathcal{Q} \Psi=0$ in the BPS background, 
the linearized version of the bosonic equations of motion \eqref{eq:bosonic_eom} 
can be obtained by expanding $\delta \Psi$ in terms of the bosonic fluctuations. 
It is convenient to impose the following gauge fixing condition 
for the bosonic fluctuations $(\delta A_{\xi}, \delta A_{\bar{z}}, \delta \phi)$:
\beq
\left( \p_z \delta A_{\bar{z}} + \D_{\bar{\xi}} \delta A_{\xi} - \frac{i}{2} \delta \phi \bar{\phi} \right) + (c.c.)  = 0.
\label{eq:gauge-condition}
\eeq
Then, $\mathcal{Q} \Psi$ can be expanded as
\beq
\mathcal{Q} \Psi
~\approx~
2(\nabla - \D_{\xi} - \D_{\bar{\xi}}) \Phi, \hs{10}
\Phi ~\equiv~
{\renewcommand\arraystretch{1.1}
\ba{c} 
- \delta A_{\xi} \\
\phantom{-} \delta A_{\bar{z}} \\
\frac{-i}{\sqrt{2}} \delta \phi \\
0
\ea},
\label{eq:linear}
\eeq
where $\approx$ denotes the equality up to the first order in the fluctuation. 
Note that $\delta \Psi \approx 0$ is the linearized version of the BPS equations \eqref{eq:BPS1} and \eqref{eq:BPS2} . 
Eqs.\,\eqref{eq:bosonic_eom} and \eqref{eq:linear} imply that 
the linearized bosonic equations of motion is $\nabla (\nabla - \D_{\xi} - \D_{\bar{\xi}}) \Phi = 0$ 
and hence
\beq
\Delta_B = \nabla (\nabla - \D_{\xi} - \D_{\bar{\xi}}).
\eeq

Thus, the eigenequations for the fluctuations are given by
\beq
\nabla (\nabla -\D_{\xi} - \D_{\bar{\xi}} ) \Phi = m_B^2 \Phi, \hs{10}
\nabla \Psi = i m_F \Psi.
\eeq
Note that $\Phi$ has to satisfy
\beq
P \Phi = \Phi , \hs{10} P = {\rm diag} (1,\,1,\,1,\,0).
\eeq

\subsection{Supermultiplets}
Since the background BPS configuration preserve the supersymmetry, 
the bosonic and fermionic fluctuations form supermultiplets. 
We can find maps between bosonic and fermionic eigenmodes as follows.

\paragraph{Fermionic Eigenmode to Bosonic Eigenmode \\}
Since the background field satisfy the BPS equation \eqref{eq:BPS2}, 
the operator $\nabla$ commutes with $\D_{\xi}$ and $\D_{\bar{\xi}}$
\beq
[\nabla, \D_{\xi}] = [ \nabla, \D_{\bar{\xi}} ] = 0,
\eeq
and hence they can have simultaneous eigenvectors. 
Assume that $\Psi$ is a fermionic eigenmode such that
\beq
\nabla \Psi = i m_F \Psi, \hs{10} \D_{\xi} \Psi = i m_{\xi} \Psi, \hs{10} \D_{\bar{\xi}} \Psi = i m_{\bar{\xi}} \Psi.
\eeq
Define $\Phi$ by
\beq
\Phi = P \Psi.
\label{eq:map-fb}
\eeq
Since $\Delta_B \Psi = m_F(m_{\xi} + m_{\bar{\xi}} - m_F)$ and $[P,\Delta_B] =0$, 
$\Phi$ is a bosonic eigenmode with eigenvalue $m_B^2 = m_F (m_{\xi} + m_{\bar{\xi}} - m_F)$. 

\paragraph{Bosonic Eigenmode to Fermionic Eigenmodes \\}
Assume that $\Phi$ is a bosonic eigenmode with eigenvalues 
\beq
\nabla (\nabla - \D_{\xi} - \D_{\bar{\xi}}) \Phi = m_B^2 \Phi, \hs{5} 
\D_{\xi} \Phi = i m_{\xi} \Phi, \hs{5}
\D_{\bar{\xi}} \Phi = i m_{\bar \xi} \Phi.
\eeq
Define $\Psi$ by
\beq
\Psi = (\nabla - \D_{\xi} - \D_{\bar{\xi}} + i m ) \Phi. 
\label{eq:map-bf}
\eeq
This is a fermionic eigenmode with eigenvalue $m_F=m$
if $m$ satisfies $m_B^2 = m (m_\xi + m_{\bar{\xi}} - m)$, i.e.,
\beq
m ~=~ \frac{1}{2} \bigg[ (m_{\xi} + m_{\bar{\xi}}) \pm \sqrt{(m_{\xi} + m_{\bar{\xi}})^2 - 4m_B^2} \bigg] \hs{5} (\equiv m_{\pm}).
\eeq
Thus, we can find two fermionic eigenmodes with eigenvalues $m_{\pm}$ 
for each generic bosonic eigenmode.

Using the maps Eqs.\,\eqref{eq:map-fb} and \eqref{eq:map-bf}, 
we can find several types of supermultiplets. 
Let us start with the bosonic component. 
Since $[P \nabla, \Delta_B] \Phi = 0$, 
we can assume that $\Phi$ is an eigenfunction of $P \nabla$:
\beq
P \nabla \Phi = i m \Phi. 
\eeq

\paragraph{Generic eigenmodes and ghosts \\}
For a generic eigenmode with $m \not = m_{\bar{\xi}}$, 
$\Phi$ takes the form
\beq
\Phi = 
{\renewcommand\arraystretch{1.1}
\ba{c} 
- \delta A_{\xi} \\
\phantom{-} \delta A_{\bar{z}} \\
\frac{-i}{\sqrt{2}} \delta \phi \\
0
\ea},  \hs{10}
\delta A_{\xi} = i (m_{\bar{\xi}}-m) \alpha, \hs{5} 
\delta A_{\bar{z}} = - \p_{\bar{z}} \alpha, \hs{10} 
\delta \phi = i \alpha \phi , 
\eeq
where $\alpha$ is a function satisfying
\beq
\left( - \p_z \p_{\bar{z}} + \frac{1}{2} |\phi|^2 \right) \alpha =  |m-m_{\xi}|^2 \alpha. 
\eeq
This is actually an eigenmode of $\nabla$ with eigenvalue $m$. 
We can show that contributions from these modes are canceled by 
those from the ghosts for the gauge fixing condition Eq.\,\eqref{eq:gauge-condition}. 

\paragraph{Long multiplets \\} 
From the explicit form of $\nabla$, 
we can show that $\delta A_{\xi} = 0$ 
for any eigenmode of $P \nabla$ with eigenvalue $m = m_{\bar{\xi}}$. 
Then, it follows that the map Eq.\,\eqref{eq:map-bf} gives two fermionic eigenmodes $\Psi_{\pm}$ 
such that $P \Psi_{\pm} \propto \Phi$. 
Thus, a generic supermultiplet of this type consists of one bosonic and two fermionic eigenmodes. 
Since the eigenvalues are related by $m_B = m_{+} m_{-}$, 
these multiplets do not contribute to the ratio of the one-loop determinants in Eq.\,\eqref{eq:saddle_point}.

\paragraph{Short Multiplets \\}
The supermultiplets which have non-trivial contribution are 
short multiplets which can be obtained by requiring that 
$\Phi$ is an eigenmode of $\nabla$ with eigenvalue $m = m_{\bar{\xi}}$: 
\beq
\nabla \Phi = i m_{\bar{\xi}} \Phi.
\label{eq:eigen-nabla}
\eeq
This is a bosonic eigenmode with eigenvalue $\Delta_B = m_{\xi} m_{\bar{\xi}}$ 
and the map Eq.\,\eqref{eq:map-bf} gives only one partner fermionic mode $\Psi \propto \Phi$ 
with eigenvalue $\Delta_F= m_{\bar{\xi}}$. 
Thus the ratio of the determinant reduces to
\beq
\frac{\det \Delta_B}{\det \Delta_F} = \prod_{\rm short} \frac{1}{m_{\xi}} = \frac{1}{\det (- i \D_\xi)} \bigg|_{\rm short}.
\eeq 

Since $\delta A_{\xi} =0$, 
the eigenmode equation \eqref{eq:eigen-nabla} reduces to 
\beq
{\renewcommand\arraystretch{1.1}
\ba{cc} 
\p_z & \frac{\bar{\phi}}{\sqrt{2}} \\
- \frac{\bar{\phi}}{\sqrt{2}} & - \D_{\bar{z}} 
\ea 
\ba{c} \delta A_{\bar{z}} \\ \frac{-i}{\sqrt{2}} \delta \phi \ea} = 0.
\eeq
This is the linearized version of the BPS vortex equation \eqref{eq:BPS1}, 
whose solution can be formally written as
\beq
\delta A_{\bar{z}} = - \frac{i}{2} \p_{\bar{z}} \delta \omega, \hs{10} 
\delta \phi = \sqrt{r} \, e^{-\frac{1}{2} \omega} \left[ \delta h - \frac{1}{2} \delta \omega h(z) \right] ,
\label{eq:eigenmodes}
\eeq
where $\omega$ and $h(z)$ are the function appearing in the background solution Eq.\,\eqref{eq:votex_sol}
and $\delta \omega$ is the function satisfying
\beq
\left( - \p_z \p_{\bar{z}} + \frac{1}{2} |\phi|^2 \right) \delta \omega = e^{-\frac{1}{2} \omega} \bar{\phi} \, \delta h(z).
\label{eq:linear-master}
\eeq
This solution is a eigenfunction of $\D_\xi$, 
if one of the holomorphic functions 
$\delta h(z)=(\delta h_1, \, \delta h_2, \, \cdots, \delta h_N)$ is non-zero and takes the form
\beq
\delta h_b(z) = z^l \exp i \left( \frac{n_3 x_3}{R_3} + \frac{n_4 x_4}{R_4} \right). 
\eeq
For such a eigenmode, the eigenvalue $m_{\xi}$ is given by
\beq
m_{\xi} = - \frac{i}{2} \left( \frac{n_3}{R_3} - i \frac{n_4}{R_4} \right) + (m_{b} - m_{a}) + ( l - k ) \epsilon,
\eeq
where we have assumed $\phi_{a} \propto z^k$ in the BPS background.

\subsection{Normalizability}\label{appendix:normalizability}
In the above discussion, 
we did not take into account the normalizability of the eigenmodes. 
The eigenmodes Eq.\,\eqref{eq:eigenmodes} with $l \geq k$ are non-normalizable 
since the boundary condition $\omega \rightarrow \log |z|^{2k}$ implies that
\beq
\sqrt{r} \, e^{-\frac{1}{2}\omega} \delta h(z) ~\approx~ \sqrt{r} \, z^l/|z|^k \rightarrow \infty \hs{5} \mbox{(for large $|z|$)}.
\eeq
We can solve this problem by making all the modes normalizable in the following way. 
Since the partition function is invariant under any deformation of $\mathcal{Q} V$, 
we can freely change the normalization factor of the kinetic terms in $\mathcal{Q} V$.
Let us consider the following deformation of $V$
\beq
V &=& \frac{1}{e^2} \left( \lambda_z \overline{\delta \lambda_z} + \lambda_0 \overline{\delta \lambda_0} + \bar{\lambda}_0 \overline{\delta \bar{\lambda}_0} + \bar{\lambda}_{\bar{z}} \overline{\delta \bar{\lambda}_{\bar{z}}} \right) \notag \\
&+& e^{-\alpha} \Big[ \psi_0^a \overline{\delta \psi_0^a} + \psi_{\bar{z}}^a \overline{\delta \psi_{\bar{z}}^a} + \bar{\psi}_z^a \overline{\delta \bar{\psi}_z^a} + \bar{\psi}_0^a \overline{\delta \bar{\psi}_0^a} + i (\lambda_0 + \bar{\lambda}_0) (|\phi_a|^2 - r) \Big], \phantom{\frac{1}{2}}
\eeq
where $\alpha$ is a function of $|z|^2$ such that $\alpha \rightarrow \infty$ for large $|z|$. 
By the field redefinition
\beq
\phi = e^{\frac{1}{2}\alpha} \hat{\phi}, \hs{5} 
\psi = e^{\frac{1}{2}\alpha} \hat{\psi}, \hs{5} 
A_i dx^i = \hat{A}_i dx^i + \frac{d \alpha \ }{d |z|^2} ( x_1 e_2 - x_2 e_1 ),
\eeq
the action takes the original form with the following position-dependent FI parameter
\beq
\hat{r} = r e^{-\alpha} + \frac{2}{e^2} \p_{z} \p_{\bar{z}} \alpha.
\eeq
Since boundary condition for the profile function $\omega$ does not change, 
the asymptotic form of the eigenmode becomes
\beq
\sqrt{\hat{r}} \, e^{-\frac{1}{2} \omega} \delta h(z) ~\approx~ \sqrt{r} \, e^{-\frac{1}{2} \alpha} z^l/|z|^k \rightarrow 0 \hs{5} \mbox{(for large $|z|$)},
\eeq
and hence all the eigenmodes become normalizable 
if the function $\alpha$ is chosen so that $e^{-\frac{1}{2} \alpha}$ decreases faster than polynomials.
We can also show that this deformation eliminates all the unpaired fermionic eigenmodes
which can exist when there are negatively charged chiral multiplets (e.g. anti-fundamental fields).

\bibliographystyle{ytphys}

\bibliography{vortex}
\end{document}